\documentclass[draft.prb,nofootinbib]{revtex4}
\usepackage{graphicx}

\newcommand\ie {{\it i.e. }}
\newcommand\eg {{\it e.g. }}

\newcommand\etal{{\it et. al. }}
\newcommand\grad{\vec\nabla}
\newcommand\half{\frac 1 2 }

\newcommand{\pa}[1]{\partial_{#1}}
\newcommand{\pref}[1]{(\ref{#1})}

\newcommand\noi{\noindent}

\newcommand{\ee}{\end{eqnarray}}

\newcommand {\be}[1]{
\begin{eqnarray} \mbox{$\label{#1}$} }

\newcommand{\rhob}{\overline\rho}
\newcommand{\too}{topological order}

\renewcommand {\r}{\vec r}
\newcommand{\tply}{topologically\ }
\newcommand{\ahm}{abelian Higgs model}
\newcommand{\pla}{ {\cal P} }

\begin{document}
\title{Superconductors are topologically ordered}
\author{T. H. Hansson$^{1,2}$, Vadim Oganesyan$^{1}$ and S. L. Sondhi$^{1}$}
\affiliation{$^{1,}$Department of Physics, Princeton University,
Princeton, NJ 08544, USA  \\
 $^{2}$ Fysikum, Stockholm University, AlbaNova,
SE-106 91 Stockholm, Sweden }

\date{\today}

\begin{abstract}
We revisit a venerable question: what is the nature of the ordering
in a superconductor? We find that the answer is properly that
the superconducting state exhibits topological order in the sense of
Wen, \ie that while it lacks a local order parameter, it is sensitive
to the global topology of the underlying manifold and exhibits an
associated fractionalization of quantum numbers. We show that
this perspective unifies a number of previous observations on
superconductors and their low lying excitations and that this complex
can be elegantly summarized in a purely topological action
of the ``$BF$'' type and its elementary quantization.
On manifolds with boundaries, the $BF$ action correctly predicts
non-chiral edge states, gapped in general, but crucial for fractionalization
and establishing the ground state degeneracy. In all of this
the role of the physical electromagnetic fields is central.
We also observe that the $BF$ action describes the topological order
in several other physically distinct systems thus providing an
example of topological universality.

\end{abstract}
\pacs{PACS numbers: xxx, xxx, xxx}
\maketitle

\section{Introduction}

\subsection{Generalities}

The notions of order and disorder are fundamental to modern condensed
matter physics. In their most influential form, starting with Landau
and now covered in textbooks \cite{chluben}, they involve ordering as
the breaking of a symmetry characterized by a non-zero local order
parameter which is the expectation value of a (generally tensor) local
operator,
\be{}
\psi(\r) = \langle \hat{\psi}(\r) \rangle
\ee
and disorder as the lack of such a broken symmetry,
\be{}
\psi(\r) = 0 \ .
\ee

Disordered states include classical gases and liquids,
paramagnets, the Bose gas above condensation, and the Fermi liquid. The
 study of their instabilities to the much more numerous broken symmetry
states has been an immensely fruitful endeavor, as reflected \eg
in the variety of Fermi surface instabilities that signal the
onset of order in fermion systems. Ordered states such as Neel
antiferromagnets, superfluids and the forest of liquid crystal
phases exhibit a rich set of interlinked properties that follow
from the broken symmetry: Goldstone bosons, topological defects
connected to dissipation, generalized rigidity and long range
forces due to the rigidity \cite{anderson84}. All of these are
captured elegantly in the mathematics of the sigma-model
Lagrangian, \be{} {\cal L} =\frac { \rho_s} 2 [\nabla
\theta(\r)]^2 \ee where the field $\theta(\r)$ contains all
fluctuations of $\psi(\r)$ with its amplitude frozen\footnote{Of
course in a continuum description, the amplitude must go to zero
on some lower dimensional manifold at the positions of the
topological defects.}.

An important theme in current research in quantum condensed matter
physics, specifically in the study of strongly correlated systems,
is the examination of systems where this framework fails to apply.
The breakdown of the framework is interesting on both sides of the
dichotomy. Are there disordered states that fail to be
characterized by their lack of a local order parameter, \ie  are
not adiabatically connected to the canonical disordered states?
Are there ordered states that also fail to be sufficiently
characterized by the order parameters they do develop? In both
cases, a related but distinct question is the existence of states
with fractionalized quasiparticles which must therefore
necessarily fail the test of continuity. A hybrid possibility, of
great interest in the context of the cuprates, is that of
accessing conventional broken symmetry states from unconventional
disordered states---in this fashion circumventing the standard
limitations on the strength of the ordering as well as on the
competitiveness of various instabilities\footnote{In the cuprates
there is evidence that the state above $T_c$ is anomalous but also
that the superconducting state is continuously connected to the
BCS state. The presence of more than one order parameter in
regions of their phase diagram, as shown recently in a set of
experiments \cite{compete}, raises the possibility that there are
competing instabilities of the high temperature state. In the
conventional Fermi liquid analysis at weak coupling, one generally
finds that one instability dominates over all the others so the
prospect of getting the competition from a non-Fermi liquid normal
state is attractive.}.

In this context the notion of ``topological order'' first articulated by Wen
and collaborators in their
studies of the quantum Hall states and the hypothesized chiral
spin liquids, is especially important \cite{wencsl,wen&niu,wen91,wenrev}.
In these instances, the states lack
local order parameters but display a weak form of order in which they
are  sensitive to the topology of the underlying two dimensional
manifolds.
Most strikingly they exhibit fractionalized quasiparticles. Further, all of
these properties are encapsulated in purely topological, Chern-Simons actions
that play a role analogous to the sigma model in broken symmetry states.

While topological order has been generally invoked in discussions of various
exotic states, our contention in this paper is that it is, in fact, the
proper characterization of the familiar superconducting state discovered by
Kammerlingh Onnes. Indeed, we find that this point has been made early on,
albeit without elaboration, by Wen himself \cite{wen91}. In this paper we will
offer a fairly complete treatment of this idea. Before turning to a more precise
statement of this
claim, we digress to list the set of properties a topologically ordered state
can be expected to exhibit by appealing to the example of the
$\nu=1/3$ fractional
quantum Hall state.

\subsection{Topological Order in Quantum Hall States}

As an instance of a topologically ordered state, the  $\nu=1/3$ fractional quantum
Hall state exhibits the following relevant properties.

\begin{itemize}

\item It does not develop a local order parameter, \ie all operators
constructed from finite numbers of electron operators exhibit
exponentially decaying correlations. It {\it does} develop a non-local,
infinite particle, order parameter but as we shall discuss later this feature
is not common to all \tply ordered systems.
When the problem is exactly reformulated as that of
a matter field coupled to a Chern-Simons gauge field,
there is no local gauge invariant order parameter \cite{gmop}.

\item Nevertheless, the system is sensitive to the topology of the
underlying manifold. It exhibits a ground state multiplet on finite systems,
separated from other states by an amount parametrically larger than the
intra-multiplet splitting, whose degeneracy increases with the genus, $g$,
of the manifold as $3^g$, \eg three states on the torus (Fig.~\ref{f1}).

\begin{figure}
\includegraphics[width=5in]{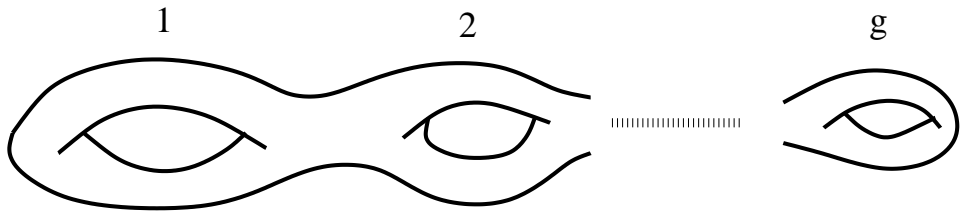}
\caption{The $\nu=\frac{1}{3}$ Laughlin liquid lacks a local order parameter, but is sensitive to the topology - on a surface of genus $g$ it exhibits $3^g$ ground states.}
\label{f1}
\end{figure}

\item The state supports fractionalized quasiholes and
quasielectrons with charge $\pm e/3$ which exhibit fractional braiding
statistics in which they acquire a phase $e^{\pm i \pi/3}$ upon
exchange among or between themselves. The existence of these
quasiparticles is intimately related to the intra-multiplet splitting
of the ground states. Their tunneling around non-contractible loops
moves the system around in the ground state manifold and leads to
the characteristic $O(e^{-L})$ ground state splitting in a
generic system of finite linear dimension $L$. In the special case
of the $\nu=1/3$ state at exactly that filling in a clean
system, the splitting vanishes altogether.

\item In the clean case one can identify a topological symmetry
algebra containing operators that move the system between different
members of the ground state multiplet. These operators insert flux
through the various non contractible loops.

\item All of the above can be encoded in a long wavelength, purely
topological, Chern-Simons Lagrangian,
\be{l-cs}
{\cal L} = {k \over 4 \pi}
\epsilon^{\mu \nu \sigma} a_\mu \partial_\nu a_\sigma - j_{\mu}a^{\mu}
\ee
with $k=3$. The elementary quantization of this action
defines
a theory with a finite dimensional Hilbert space with the proper
ground state degeneracies and its correlations in the presence of sources
reproduce the quantum numbers of the quasiparticles.

\item The topological action further implies the existence of
boundary degrees of freedom on manifolds with boundaries.
In the case of the Laughlin quantum Hall states, the boundary excitations
form a chiral Luttinger liquid.

\end{itemize}

\subsection{This paper}

In this paper we will show that superconductors with a gap in their single
particle spectrum exhibit appropriate versions of all of the above
properties: lack of a local order parameter, topological degeneracies and
symmetry algebra,\footnote{We should note that the topological symmetry
operators are not expected to be universal everywhere in a topologically
ordered phase as shown by example in Ref.~\onlinecite{msf2001}.}
fractionalization, description
by a topological field
theory and edge degrees of freedom, and hence are properly described as
being topologically ordered. In this discussion it will be crucial that
superconductors are not mere superfluids like $^3$He and $^4$He
but instead are {\it charged superfluids with dynamic electromagnetism}.

The claim of topological order for superconductors might surprise
some fraction of our readers on at least two of its component
pieces---that superconductors are not broken symmetry states and
that they exhibit quantum number fractionalization. In fact, both
ideas have been around for a while. The impossibility of finding a
gauge invariant local order parameter for the state {\it in the
presence of electromagnetic gauge fields} has been understood for
a long time \cite{elitzur}. The conventional broken symmetry
account, following Bardeen, Cooper and Schrieffer, holds for a
neutral system whose response functions are argued to be
qualitatively similar to the ``screened'' or ``irreducible''
response functions of the charged system. The point that the
quasiparticles of a superconductor are electrically neutral, and
hence fractionalized, was made (only!) a decade back by Rokhsar
and Kivelson and again they invoked the gauge field in an
essential way\cite{kiv90}.

In the following we will be able to add to these observations an
account of degeneracies on closed manifolds, a topological action, and an account
of the edge states it predicts, to produce a unified portrait of
topological order which can then substitute for the lack of a broken symmetry. As
befits a topic with an extensive scholarly literature, we have found that
much of what we have to say has precursors in the literature which we note
at various points in the text.
A subsidiary theme in this paper is that more than one system can exhibit
the same
topological structure and hence be described by the same topological field
theory, and we will find it instructive to examine the correspondences. In
particular we will note that the standard Ising gauge theory, the short ranged
RVB state, a bilayer quantum Hall system with oppositely charged layers,
and a $U(1)$ lattice gauge theory coupled to a charge-2 scalar, will
give rise to the same topological structure as the superconductor. Indeed,
the last one on that list, studied in the seminal work of Fradkin and
Shenker \cite{frad79}, illustrates our central points very elegantly.

In our discussion we will largely shy away from truly microscopic models of
the superconducting state with the electronic degrees of freedom
exhibited explicitly, since that level of detail is not essential for
our considerations. Instead we shall study bosonic theories of the quantum
Ginzburg-Landau form. In field theoretic terminology these are the
relativistic abelian Higgs models governed by the Lagrangian,
\be{abh}
{\cal L}_{ah} = \frac 1 {2M} |iD_{\mu}\phi|^{2} -
\frac \lambda 4 (\phi^\dagger\phi)^{2} - \frac {m^2} 2 \phi^\dagger\phi
- \frac 1 4 F_{\mu\nu}^{2} - eA_{\mu}j^{\mu} \ .
\ee
Here $\phi$ is a charge $-2e$ scalar field representing the
Cooper pair condensate, the covariant derivative
$iD_{\mu} = i\partial_{\mu} -2eA_{\mu}$ and the field strength
$F_{\mu\nu}$ involve the physical electromagnetic field and the conserved
current $j^{\mu}$ with charge $e$
is introduced to describe the gapped quasi-particles
or perhaps external charges \footnote{
Note that in spite of the relativistic form we
  normalize the kinetic term such that $|\phi|^2$ has the
  dimension of density. This will help to
  streamline our notation with
  that usually used in discussing superconductivity.
  In the non-relativistic limit this model becomes
  a time dependent Ginzburg-Landau theory. This model would
  exhibit the Meissner effect with a London penetration length
  $\lambda_L$ coming from the gradient term. The Debye
  screening due to the Coulomb field would, however, only
  be generated because of the scalar
  potential, and the corresponding
  screening length would be given by $\lambda_D \neq \lambda_L $.
  In the relativistic model both electric and magnetic screening
  emanate from the gradient term, and the two screening lengths are
  equal. Although this is not true in real systems,
  it simplifies our arguments and helps to highlight the conceptual
  points. The generalization to a real non-relativistic
  model is left for the reader.}
(We will use Greek and Roman indices to denote space-time vectors and
spatial vectors respectively, and the metric $g^{00}=1$ and
$g^{ii}=-1$.)
In 3+1 dimensions, this model is a plausible description of a gapped BCS
superconductor with particle-hole symmetry  but
it has the topological features of interest even if the choice of a Lorentz
invariant dynamics is non-generic. As an aside we note that
the situation is more complicated
for gapless superconductors, \eg the d-wave cuprates, where there
are gapless quasi-particles that must be incorporated in the effective
low energy theory. We will return to this in the summary section.

In a final simplification, we will focus mostly on ${\cal L}_{ah}$
in 2+1 dimensions. This no longer describes a physical
superconductor since the electrodynamics is now that of the 2+1
dimensional Maxwell term (for instance a logarithmic potential
between charges) which  does not describe real electromagnetism
even if the electron system is effectively two dimensional as is
the case with superconducting films.\footnote{ The results
presented in this paper are probably valid for thin charged
superconducting films anyway. In this case we have power law
rather than exponential decay of screening charges and currents,
which appears sufficient to define a appropriate scaling limit and
thus allow for a description in terms of a topological field
theory.} The primary reason to examine this case nevertheless is
that the analysis is simpler and more pedagogical than in 3+1
dimensions while the essential features of the two problems are
the same. The chief simplification is that the topological theory
for a 3+1 dimensional superconductor is a theory of particles and
strings, while for 2+1 dimensions it is theory of particles only.
Further, on manifolds with boundaries, the boundary theories of
the 2+1 dimensional models are 1+1 dimensional, which are even
easier to discuss. A secondary reason is that various theories of
strong correlation in 2+1 dimensions give rise to the identical
mathematics of coupled matter and gauge fields for physically {\it
neutral} systems and our discussion will serve to formalize the
discussion of topological order in that context as well. We should
emphasize though, that while the occurrence of \too\ in this class
of theories is a fascinating question, especially with regard to
the physics of the non-superconducting regions of the cuprate
phase diagram, it has nothing to do with the \too\ in the
superconducting phase itself. In all such models, the real
electromagnetic field would eventually be important to establish
the topological order of the 3 dimensional superconducting
state---a statement which should be self-explanatory at the end of
the paper.

With this somewhat elaborate preamble we now turn to the technical content
of the paper. In the next section, we briefly review why a
superconductor cannot be characterized by a broken symmetry, \ie
why there is no gauge invariant local order parameter. In section III
we discuss the nature of the excitations in a charged superconductor,
and why they are fractionalized.
From these we deduce the form of the topological
$BF$ action, which we then rederive
from a path integral formulation of the abelian Higgs model.
This action implies a ground state degeneracy which we discuss in Section IV.
In Section V we digress to consider other problems that are also described by
the $BF$ theory: the lattice $Z_2$ gauge theory, a bilayer quantum Hall
system, the resonating valence bond (RVB) state and the Fradkin-Shenker
problem.
In Section VI we turn to the edge structure implied by the $BF$ action
in 2+1 dimensions as well as in 3+1 dimensions.
The last section summarizes our main results and some technical details
connected to edge actions are in an appendix.

\section {No local order parameter}

The textbook Ginzburg-Landau description of a gapped
superconductor invokes a charge $-2e$ complex scalar
field, the ``superconducting order parameter'', that
measures the condensation of Cooper pairs and is related
to the underlying electron field by an appropriate
expectation value, $\psi(\vec r) = \langle \Psi_\uparrow
(\vec r) \Psi_\downarrow (\vec r) \rangle$. This field is
minimally coupled to the electromagnetic vector
potential $A^\mu$ and the dynamics of the two coupled fields
are then fixed by the Ginzburg-Landau differential
equations.  These equations are, obviously, a fine
description of superconductors with small fluctuations.
At issue in the context of this paper is whether $\psi(\vec r)$
is a local quantity.

To see that it is not, let us rephrase the question in the
context of the \ahm \, (\ref{abh}). The Euler-Lagrange
equations for ${\cal L}_{ah}$ absent sources are of the
Ginzburg-Landau form, although now extended to include
a precessional dynamics at $T=0$. We expect the Euler-Lagrange
equations to give a useful account if the fluctuations are
small in the ordered phase and the fields involved develop
non-zero expectation values. Naively, we would like $\phi$
to develop a nonzero expectation value but this is not
possible since it transforms non-trivially under the
$U(1)$ gauge symmetry,
\be{gt}
\phi (\vec r) \rightarrow e^{i2e \alpha (\vec r)} \phi (\vec r) \ \ \ \ \ \ \
; \ \ \ \ \ \
A_\mu (\vec r) \rightarrow
A_\mu (\vec r) + \partial_\mu \alpha (\vec r) \nonumber \, ,
\ee
and Elitzur's theorem \cite{elitzur} assures us that such
quantities average to zero even in the ``broken symmetry''
phase.

The solution to the conundrum of what underlies the
Ginzburg-Landau description is the {\it non-local} quantity first
introduced by Dirac\cite{dirac}. It is easiest to write it in
operator form, \be{dirac} \phi^\dagger_D(\vec r ) = e^{i\int
d^3r'\, \vec E_{cl}(\vec r'- \vec r)\cdot \vec A(\vec r')}
\phi^\dagger(\vec r ) \ee where $\vec E_{cl}(\vec r)$ is the
classical electric field corresponding to a point charge at the
origin, \ie $\vec\nabla\cdot\vec E_{cl} = \delta (\vec r)$, and
$\phi$ and $\vec A$ are quantum field operators. A partial
integration shows that the gauge transformation \pref{gt} leaves
the combination $\phi_D$ invariant. The operator $\phi_D$ has a
natural interpretation as the creation operator of a charged
$\phi$ particle together with a coherent states of photons
describing the accompanying Coulomb field which extends out to
infinity. In the Coulomb gauge $\grad\cdot\vec A = 0$, the Coulomb
field is described entirely by the scalar potential, $A^0$, and
$\phi_D$ reduces to  $\phi$ alone. So in this gauge the Dirac
order parameter {\it appears} local, as can be checked by writing
$\vec E_{cl}(\vec r)$ as a gradient and again integrating by
parts. A superconductor is then characterized by off-diagonal long
range order in $\phi_D$. Kennedy and King have given a rigorous
proof of this statement using a covariant generalization of
\pref{dirac}, and a lattice regularization, for a non-compact
abelian Higgs model in two or more spatial dimensions\cite{kk85}.

Their proof also shows that this non-local order parameter cannot be
used as one uses a local order parameter. Precisely, one finds that
the temporal correlator of $\phi_D$ decays {\it algebraically} to
its asymptotic value.  With a local order parameter this would be
a signature of Goldstone bosons. In fact, the Anderson-Higgs mechanism
forbids any such bosons in the actual spectrum, which shows that a
description based on $\phi_D$ does not have the character of
the standard sigma model.

While we are on the subject of non-local characterizations of the
superconductor, a second possibility, following 't Hooft, is to
classify phases by focussing on observables inspired by the
behaviour of the gauge sector.\footnote{A similar construction,
but for gauge fields alone, was given earlier by Fradkin and
Susskind \cite{fradsussdual}.} Here the candidates are Wilson
loops, and their duals, which correspond mathematically to the
insertion of singular gauge transformations\cite{thooft78}.
Physically, these dual variables ask a dimension dependent
question. In 3+1 dimensions, in a superconducting phase with
non-compact electrodynamics, 't Hooft's operator is a loop whose
area law decay attests to the confinement of test magnetic
monopoles by the Abrikosov flux tube that gets stretched between
them. In 2+1 dimensions the t'Hooft operator $\phi_m$ acts at a
point and becomes a nominally local field, $\phi_m$ creating a
vortex. This yields a disorder parameter, which vanishes in the
superconductor and has a finite expectation value in the normal
phase of the \ahm. In words, the normal phase is identified as a
condensate of vortices while the superconductor exhibits a gap to
their creation.

In both of the above characterizations the restriction to
non-compact gauge fields is not accidental. In a compact 3+1
dimensional gauge theory there are monopoles that obstruct the
construction of $\phi_D$ and its covariant generalizations so that
even a nonlocal order parameter in the spirit of Dirac is not
possible\cite{fm01}. The essential difficulty is that the Dirac
quantization condition is not compatible with having a real valued
current as in \pref{dirac}. It is even easier to see what goes
wrong with the t'Hooft construction once dynamical monopoles are
permitted. For example in 3+1 dimensions, without them, the
potential energy of two static test monopoles separated by a
distance $r$ in a superconductor will be linear $\sim \sigma r$
where $\sigma$ is the energy per unit length \ie the tension of
the Abrikosov flux line. In the presence of dynamical monopoles of
mass M, the linear confinement will breakdown at a distance
$r_{sc} \sim 2M/\sigma$ where it will be energetically preferable
to create a monopole-antimonopole pair from the vacuum to break up
the flux line. This is the exact magnetic analog of  electric
screening of static electric test charges in a confining
relativistic theory with massive charged particles.

While this discussion will certainly be germane when we discuss
some compact gauge problems related to our main theme, readers
interested solely in superconductors may suspect that they can
do without it altogether. While this is true in practice, it
is probably not true as a matter of principle! While Maxwell
electrodynamics and indeed even the standard model have no
monopoles, they do occur in most attempts at further unification,
\eg in various grand unified models, with masses expected to be
in the $10^{15} - 10^{16}$ GeV range\cite{weinbII}. With such
masses they will give rise to a ``screening length'' that
we can estimate, for superconductors with Ginzburg-Landau
parameter $\kappa = 1$ (so that the coherence length and
the penetration depth are the same), as being of the simple
form
\be{slen}
\lambda_{\rm mp} \sim \lambda_F {M c^2 \over E_F} \ .
\ee
For a good old fashioned superconductor this yields
$\lambda_{\rm monopole} \approx 10^{10} {\rm km}$ or
about $70 AU$
which is therefore literally astronomical.\footnote{
A naive estimate of the corresponding tunneling probability
based on the Schwinger formula commonly used in QCD
string phenomenology\cite{stringbreak}, gives a
string life time $\sim e^{-10^{44}}$ wherein the
units are evidently unimportant! A better estimate requires
consideration of the instanton path which we defer to the
future.}
It follows
then that for samples of this size there really won't be an
order parameter which makes it all the more imperative to develop
an alternative characterization of the order in the superconducting
state---a task to which we now turn!

\section{Excitations, Fractionalization and Topological Field Theory}

Having established that a local order parameter description is not
feasible for superconductors, we will now (successfully) attempt to
construct a topological order description in terms of a topological
field theory. We will start with the low energy excitations of the
superconducting state and examine their quantum numbers and topological
interactions. By encoding these in a topological actions in 2+1 and 3+1
dimensions we will inductively
arrive at the desired description. Subsequently we deduce the
same topological action in 2+1 dimensions
from the path integral for the \ahm\ and close
by noting that including leading irrelevant terms beyond the topological action
completes the low energy description of the superconductor, much as it
does for the quantum Hall effect.

\subsection{Excitations and Fractionalization}

The low energy excitations of a superconductor are the quasiparticles
formed by breaking up a Cooper pair, vortices or vortex lines in 2 and 3
spatial dimensions respectively, and a set of collective modes which together
form a massive photon in our relativistic setting.

To review their properties in the context of \pref{abh}, we write $\phi$ in
amplitude and phase variables, $\phi = \sqrt\rho e^{i\varphi} $ and focus
deep in the ordered phase where $m^2 \ll 0$. Here we can set
the amplitude equal to its classical value, $\rhob =- m^2/\lambda$, and
ignore its remaining massive fluctuations to rewrite \pref{abh} as
\be{abh2}
{\cal L}_{ah} = \frac \rhob {2M} (\partial_{\mu}\varphi + 2e
A_\mu)^{2}
- \frac 1 4 F_{\mu\nu}^{2} - eA_{\mu}j^{\mu} + ....
\ee
where the dots indicate the neglected density fluctuations.
If furthermore the model is defined on a topologically trivial
manifold, and we disregard vortices, we may send
$A_\mu \rightarrow A_\mu - \frac 1 {2e} \partial_{\mu}\varphi$
in a regular gauge transformation that defines unitary gauge, thus
obtaining the following effective low energy Lagrangian,
\be{eff}
{\cal L}_{eff} =
- \frac 1 4 F_{\mu\nu}^{2} + \frac {m_s^2} 2 A_{\mu}^2
-eA_{\mu}j^{\mu}
\ee
where the screening mass, $m_s$ is related to the London penetration
length by $m_s^2 = \lambda_L^{-2} = 4e^2\rhob/M $.

The Lagrangian \pref{eff} is that of a massive abelian gauge field
coupled to a conserved current. In the absence of the current it
yields the gapped collective modes of the superconductor---the absence
of a gapless mode is the Anderson-Higgs mechanism.

In the presence of a current, the classical equation of motion is a
relativistic version of the London equation,
\be{eom}
\pa\nu F^{\nu\mu} = j^\mu -  m_s^2 A^\mu = j^\mu + J^\mu_{sc} \, ,
\ee
where we identified $-m_s^2 A^\mu$ as the screening current in the
medium.  For $m_s^2\neq 0$
\pref{eom} implies $\pa\mu A^\mu = 0$, \ie the screening current is
conserved, and the equation of motion simplifies to,
\be{eom2}
(\triangle + m^2)A^\mu = j^\mu \, ,
\ee
from which it follows that {\em all} classical fields and currents
are exponentially
screened over the length $\lambda_L$. This is a consequence of the Meissner
effect and should be contrasted with the case of an
ordinary metal, where only the charge and the longitudinal part of
the current are screened. That the screening lengths for both components
are the same is special to our Lorentz invariant setting---in general,
they will be different.

This screening has important consequences for the quantum numbers
of the quasiparticles, as pointed out by Kivelson and Rokhsar
\cite{kiv90}---they do not carry a classical charge. To see this,
consider constructing a wavepacket with the quasiparticle at rest
in a given frame. In that frame the scalar potential is the only
non-zero component of $A^\mu$ and it decays to zero on the scale
of $\lambda_L$. By Lorentz transforming we obtain the potentials
for a quasiparticle in motion and still find that all components
of $A^\mu$ are exponentially attenuated beyond $\lambda_L$. As no
fields are generated beyond the screening length, the
quasiparticle is classically neutral at long wavelengths. Again we
should note that life is more complicated when the longitudinal
and transverse screening lengths are different. In the extreme
case of the metal, where the transverse screening length is
infinite, a moving charge will give rise to a dipolar pattern of
current backflow that will decay only algebraically at long
distances \cite{pnoz}. For real superconductors this dipolar
pattern will be cutoff at the scale of the London length, while
the longitudinal currents and potentials will decay on the scale
of the Thomas-Fermi length. In our problem the two parts are
screened identically and hence there is no residue of the dipolar
pattern whatsoever. This vanishing of the charge of the
quasiparticles is an instance of quantum number fractionalization
in that the fundamental electrons are charged. If the electrons
carry spin then the quasiparticles do too and hence are spinons
\cite{kiv90} but this is not central. For example, in a p-wave
superconductor of spinless fermions there would be no change in
the underlying fractionalization. Instead the proper formulation
is that {\em the quasiparticles retain a quantum, Ising charge,}
which we will discuss in the next subsection.\footnote{Readers
familiar with the work of the Santa Barbara group \cite{itp}
should note that their discussion does {\it not} involve the
physical electromagnetic field and is thus physically quite
different from that of \cite{kiv90} and ours. For us the
superconducting phase is fractionalized while in \cite{itp} it is
the non-superconducting phase that is fractionalized.}

This analysis has used the equation of motion \pref{eom} which
deals with expectation values and has sidestepped the important
question of defining operators for which the vanishing charge is a
sharp observable \cite{goldkiv}. To our knowledge, there isn't a
rigorous analysis of this question for the \ahm. There is however
a more careful account of the expectation value question by Swieca
\cite{swieca76} (for a rigorous version, see
Ref.~\onlinecite{buchholz79}). Swieca proves the following: A
theory in more than 3 space-time dimensions, with a mass gap and
an identically conserved current, \ie a current satisfying $\pa\nu
F^{\nu\mu} = j^\mu$, has no charged states in the spectrum. This
theorem is directly applicable to our model Lagrangian \pref{abh}
if we take the total current $j^\mu_{tot} = j^\mu + J^\mu_{sc}$ in
\pref{eom} as the identically conserved current. Swieca's proof,
which is based on Lorentz invariance of the current form factor,
and locality of the electromagnetic field, is not obviously
applicable to a non relativistic theory, and it would be
interesting to establish such an extension.

Finally, we note that in writing \pref{eff} we explicitly ignored
vortices and vortex loops/lines. These form the remaining low
energy excitations. A vortex carrying a flux $\pi$ is also
fractionalized in a sense that is sharpest for models with a
lattice electrodynamics as they exhibit vortices with $2 \pi$
flux as their primary excitations when decoupled from matter.

We turn now to embedding these excitations in a topological action.

\subsection{$BF$ theories }

At issue in writing down a topological action are the topological
interactions among the excitations, \ie interactions which depend
upon the topology of the field configurations (or particle
worldlines) but not on the metric. A way to formalize this is by
the idea of the topological scaling limit in which we examine
the system at scale $R$ and keep those pieces of the correlation
functions that are $O(R^0)$ as $R \rightarrow \infty$ {\it at
fixed couplings}\cite{frohlich-top}.
This limit is to be contrasted with
the Wilsonian scaling limit in which the coupling constants are
tuned so as to keep the ratio of $R$ to the correlation length
$\xi$ fixed. While the latter keeps all information except at
the lattice scale, the former keeps only the topological ``braiding''
information.

\begin{figure}
\includegraphics[width=5in]{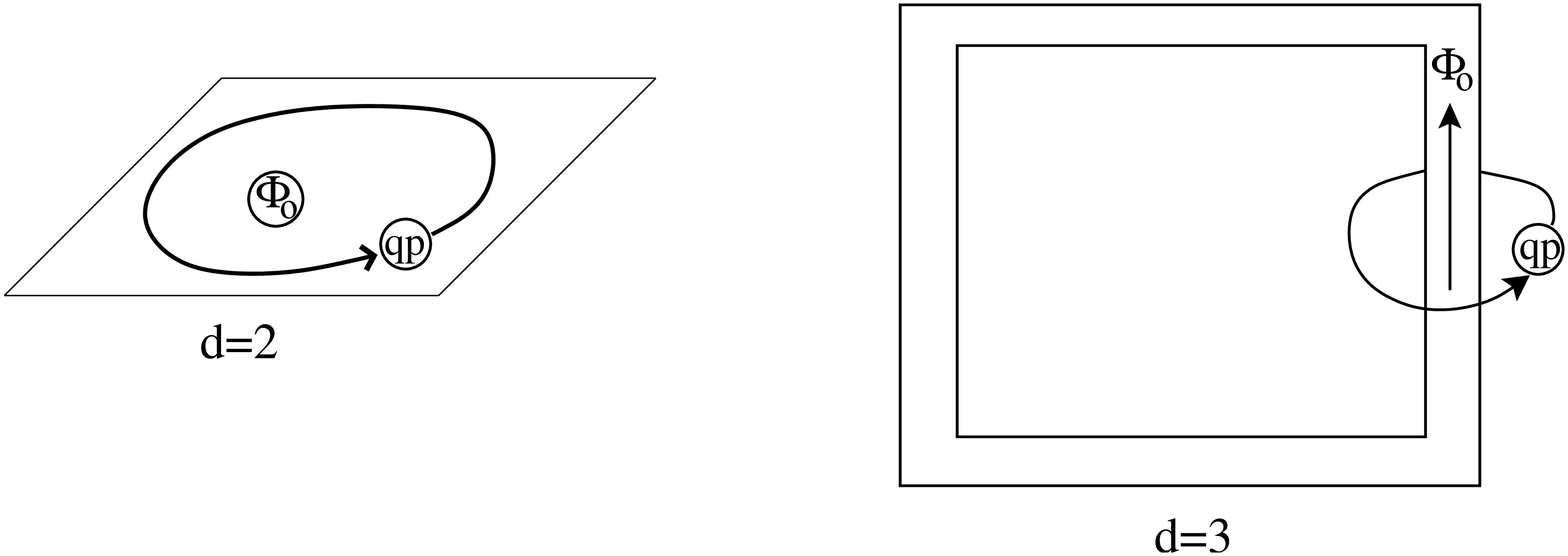}
\caption{
Topological interactions in a superconductor: quasiparticles encircling vortices
($d=2$) or threading vortex loops ($d=3$) pick up a phase $\pi$  at an
arbitrary distance.}
\label{f2}
\end{figure}

Among the quasiparticles, vortices and plasmons there is one non-trivial
interaction in this limit---namely, a topological phase $\pi$ arises
whenever a quasiparticle is transported around a vortex or vice-versa
(Fig.~\ref{f2}).
This can be read off from the venerable explicit solution of
the Bogoliubov-de Gennes equations for a vortex \cite{dGbook} but
more modern discussions of how it arises are enlightening \cite{kra89,rez89}.

The presence of this interaction is why we were
careful to refer to the {\it classical} neutrality of the quasiparticles
in the past section. Further, this interaction has the feature
that it only detects quasiparticle number modulo 2 so that quasiparticles
carry an Ising charge under it thus explaining our comment to this
effect in the last subsection. This sensitivity of the superconductor
to particle number modulo 2 has been described as an Ising gauge
invariance of the superconducting state previously \cite{kra89}.

This topological interaction can be readily written into a field theory.
We first consider the 2+1 dimensional case where both quasiparticles
and vortices are particles so we can proceed in close analogy to
the bosonic Chern-Simons theory for the quantum Hall effect and attach flux and
charge to the particles in such a way that the Berry phases
(or in the quantum Hall case, the exchange phases) appear as
an  Aharonov-Bohm effect. We define a unit charge quasiparticle
current $j^{\mu}$, and a vortex current $\tilde j^{\mu}$, and
couple them to electric and magnetic gauge potentials via
the Lagrangian,
\be{coupl}
{\cal L}_{curr} = - a_{\mu}j^{\mu} - b_{\mu}\tilde j^{\mu}
\, .
\ee
A simple calculation shows that in order to get a phase $\pi$
when moving a $j$ quantum around a $\tilde j$ quantum we need
an action for the gauge potentials, which is
of the ``$BF$'' type\cite{blau}
\be{bf}
{\cal L}_{BF} =
 \frac 1 {2\pi} \epsilon^{\mu\nu\sigma}b_\mu f^{(a)}_{\nu\sigma }
\, ,
\ee
where $f^{(a)}_{\mu\nu} = \pa\mu a_\nu - \pa\nu a_\mu$.
Putting the parts together we have the topological action,
\be{top}
{\cal L}_{top} =
\frac 1 \pi \epsilon^{\mu\nu\sigma}b_\mu\pa\nu a_\sigma
 - a_{\mu}j^{\mu} - b_{\mu}\tilde j^{\mu} \, .
\ee

The topological nature of ${\cal L}_{top}$ is clear from the
equations of motion,
\be{teom}
\tilde j^\mu &=& \frac 1 \pi
\epsilon^{\mu\nu\sigma}\pa\nu a_\sigma
= \frac 1 {2\pi} \epsilon^{\mu\nu\sigma} f^{(a)}_{\nu\sigma}   \\
j^\mu &=& -\frac 1 \pi
\epsilon^{\mu\nu\sigma}\pa\nu b_\sigma
= - \frac 1 {2\pi} \epsilon^{\mu\nu\sigma} f^{(b)}_{\nu\sigma}
\, , \label{teom2}
\ee
which show that the gauge invariant field strengths are fully
determined by the currents, just as in a Chern-Simons theory.
These equations both have a very direct physical interpretation
as we shall see later.

Two comments are in order. The first concerns the symmetry properties
of the Lagrangian \pref{top}.
Under the parity transformation $(x,y)\rightarrow (-x,y)$ the two
potentials transform
as $ (a_0,a_x,a_y) \rightarrow (a_0,-a_x,a_y)$ and
$ (b_0,b_x,b_y) \rightarrow (-b_0,b_x,-b_y)$,
while under time reversal the transformations are,
$ (a_0,a_x,a_y) \rightarrow (a_0,-a_x,-a_y)$
and $ (b_0,b_x,b_y) \rightarrow (-b_0,b_x,b_y)$, respectively.
The unusual transformation properties of the potential $b_\mu$
follows from that of
the vortex current. It is easy to check that the $BF$ action is
invariant under both $PT$ and $CPT$. Second, in the Lagrangian \pref{top}
both currents are integer valued. This quantization is naturally encoded
by requiring that the gauge fields $a_\mu$ and $b_\mu$ be compact.
In the continuum this means that they transform as
\be{compactgt}
a_i &\rightarrow& a_i + \partial_i \Lambda_a  \nonumber \\
b_i  &\rightarrow& b_i + \partial_i \Lambda_b  \ ,
\ee
with gauge functions $\Lambda_{a/b} \equiv \Lambda_{a/b} + 2 \pi$.
This compactness will also be evident in our rederivation of the
$BF$ action from the microscopic theory in the next section.

Turning to the case of 3+1 dimensions, we have essentially the
same construction, but with the difference that the vortices
are now strings, and the vector potential $b$ is  an
antisymmetric tensor, $b_{\mu\nu}$. In form language, the
action still has the structure $BF$, and written out in components
it reads\cite{blau,berg95},
\be{bf3}
{\cal L}_{BF} =
\frac 1 \pi \epsilon^{\mu\nu\sigma\lambda}
b_{\mu\nu}\pa\sigma a_\lambda \, .
\ee
The gauge transformations of the $b$ field are given by
\be{gt3}
b_{\mu\nu} \rightarrow b_{\mu\nu} + \pa\mu\xi_{\nu} -
\pa\nu\xi_{\mu}
\ee
where $\xi_{\mu}$ is a vector valued gauge parameter.
The minimal coupling of the $b$ potential to the world
sheet, $\Sigma$ of the strings is given by the action,
\be{string}
S_{vort} = - \int_{\Sigma} d\tau d\sigma^{\mu\nu}\,
b_{\mu\nu} =
- \int_{\Sigma} d\tau d\sigma\,
\left|\frac {d( x^{\mu} , x^{\nu}) }
{d(\tau , \sigma)}\right| b_{\mu\nu} \, ,
\ee
where $(\tau, \sigma)$ are time and space like coordinates
on the worldsheet. This
is a direct generalization of the
coupling of $a$ to the world line, $\Gamma$,
of a spinon,
\be{spin}
S_{sp} = -\int_{\Gamma} dx^\mu\, a_\mu
= - \int_\Gamma d\tau\, \frac {d x^{\mu}} {d\tau}
a_{\mu} \, .
\ee
Combining these elements we get the topological action for
the 3+1 dimensional superconductor,
\be{top3}
S_{top} = \int d^{4}x\, {\cal L}_{BF} + S_{sp} + S_{vort} \, .
\ee
The proof that this action indeed gives the correct
braiding phases can be found \eg in Ref.~\onlinecite{berg95}, and a
discussion of this action in the context of superconductivity
has appeared before in Ref.~\onlinecite{bala92}, more on which later.

\subsection{The 2+1 $BF$ theory from the abelian Higgs model }

Previously we induced the $BF$ action  \pref{top} from our knowledge of the
low energy excitations and their topological interactions.
We now gain additional insight into its form by deriving
it from the Lagrangian for the abelian Higgs model
\pref{abh} by explicitly including the vortices we neglected
before.

An (anti)vortex at position $\vec r$ is a solution of the
classical equations of motion where the phase, $\varphi$ of the
$\phi$ field winds $(-)2\pi $ along any closed curve encircling
$\vec r$. The generalization to higher winding numbers and to
multi-vortex configurations is obvious. For well separated points,
one can also define configurations with $N_+$ vortices and $N_-$
anti vortices, although only $N_+ - N_-$ is topologically
conserved. Away from  the vortex cores, the solutions again look
like a pure gauge, but with the important difference that the
$\pa\mu\varphi$ term  in \pref{abh2} cannot be removed by a
regular gauge transformation. Instead we split the phase field as
$\varphi = \tilde\varphi + \eta$ where $\tilde\varphi$ is a
function of the vortex positions, $\vec y_n$, and $\eta$ is the
fluctuating quantum field. We can now perform the {\em regular}
gauge transformation $A_\mu \rightarrow A_\mu -  \frac 1
{2e}\partial_{\mu}\eta$. If we consider a fixed vortex
configuration $\{ \vec y_n , q_n \}$ where $q_n = \pm 1$, we can
write the corresponding quantum partition function in terms of an
Euclidean  path integral\cite{bard78}, \be{pfunc} Z[j^\mu, \{\vec
y_n , q_n\}] = \int {\cal D} [A_\mu] {\cal D} [|\phi|] e^{-\int
d^3r\, {\cal L}_E } \ee with \be{pathl} {\cal L}_E =  \frac 1 4
F_{\mu\nu}^{2} + \frac {m_s^{2}} 2 (A_{\mu}-\frac 1 e a_\mu)^2 -
eA_{\mu}j^{\mu} +{\cal L}_{|\phi|} \, , \ee where we introduced
the notation $a_\mu = -\frac 1 2 \pa\mu\tilde\varphi$ and use the
metric $(+++)$. Here ${\cal L}_{|\phi|}$ includes both the
potential terms, density derivative terms, and an explicit
dependence on the vortex positions. The gauge field is now
manifestly massive, and with the potential in \pref{abh}, so is
the density field $|\phi|$. In the effective low energy
description the only effect of the density fluctuations that will
be retained is the presence of a vortex current, \be{vcur} {\tilde
j}^\mu (x^\mu) = \sum_n q_n\int d\tau_n \dot y^\mu \delta^2 (
x^\mu - y_n^\mu(\tau)) \, . \ee where $y_n^\mu(\tau)$ parametrize
the (Euclidean) world lines of the vortices. It will be convenient
to parametrize the vortex current with a gauge potential $b_{\mu}$
as, \be{vcur2} \tilde j^\mu (x^\mu) = \frac i \pi
\epsilon^{\mu\nu\sigma}\pa\nu a_\sigma \, \ee The normalization is
such that a unit charge $\tilde\rho = \delta^2(\vec x)$ is
associated with a fundamental vortex in the charge $-2e$ scalar
field, \ie $\int d^2x\, \tilde \rho = \frac 1 \pi \oint d\vec
x\cdot \vec a =
 \frac 1 \pi \frac 1 2 \oint d\eta = 1$.
Ignoring the density fluctuations, and hence ${ \cal L}_{|\phi|} $,
 we can now rewrite \pref{pfunc} as
\be{low}
Z[j^\mu, \tilde j^\mu ] = \int {\cal D} [A]
{\cal D}[a]\, \delta[ \tilde j^\mu -
\frac 1 \pi \epsilon^{\mu\nu\sigma}\pa\nu a_\sigma ]
e^{-\int d^3r\, ({\cal L}_E -{\cal L}_{|\phi|})}
=\int {\cal D}[a]{\cal D}[b] \, e^{-\int d^2r\,
{\cal L}_{eff}(a,b) } \, ,
\ee
where
\be{deftop}
e^{-\int d^3r\, {\cal L}_{eff} } = \int {\cal D} [A]
e^{-\int d^3r\, {\cal L}_{E} }
\ee
and
\be{llow}
{\cal L}_{E} =   \frac 1 4 F_{\mu\nu}^{2} +
\frac {\lambda_L^{-2}} 2 (A_{\mu}- \frac 1 e a_\mu)^2
- eA_{\mu}j^{\mu}
- b_\mu \tilde j^\mu + \frac i \pi \epsilon^{\mu\nu\lambda}
b_\mu \partial_\nu a_\lambda \, .
\ee
The gauge potential $b_\mu$ is a Lagrange multiplier
that imposes the delta function constraint in \pref{low}.
The remaining steps in deriving the low energy Lagrangian
$ {\cal L}_{eff} $ is to shift the field
$A_\mu  \rightarrow A_\mu + \frac 1 e a_\mu $,
perform the Gaussian integration over the massive $A$ field
and finally doing a derivative expansion.
To lowest order, and after rotating back to Minkowski
space, we get,
${\cal L}_{eff} = {\cal L}_{top} + O(m^{-2})$, \ie
the previously derived
topological action. We shall return to the higher
order corrections below.

Although this derivation was for 2+1 dimensions, essentially the
same argument can be given to derive the 3+1 dimensional action
\pref{top3}.

The physical significance of the potentials $a$ and $b$ is
now revealed: from \pref{pathl} above it is clear $a$ is
nothing but the topological part of
the usual vector potential $A$, \ie the part which is
a pure gauge everywhere except
at the location of the point vortices as expressed
by the constraint in \pref{low}.

Equation \pref{teom2} expresses screening of the external current,
since $db$ is just the dual form of the screening current $J_{sc}$
in \pref{eom}. Also from writing $\rho_{sc} =
-\epsilon^{ij}\partial_i b_j = \partial_i E^i_{sc} $ it follows
that $b_i = \epsilon_ {ij}E^j_{sc} $, \ie the potential $b$ is
essentially the  fields associated with the screening clouds
induced by the external electric sources. Since the total field is
zero, this still begs the question to how there can be any long
range effect related to the $b$ potential. Put differently, how
does a moving vortex detect a stationary charge, given that the
electric field is exponentially screened? A particularly clear
explanation has been given by Reznik and Aharonov, who showed that
although the expectation value of the electric field is
exponentially screened inside the superconductor, there is an
unscreened ``modular'' or $Z_2$ part that give rise to the
topological phase\cite{rez89}. We will return to this below in the
discussion of the ground state degeneracy.

In summary, there are three complementary ways to understand
the topological $BF$ action for the superconductor:
\begin{enumerate}
\item It encodes the correct braiding phases of charges and vortices.
\item It relates the  current of correctly normalized pointlike vortices
in the condensate to the topological nontrivial part of the
vector potential.
\item It implements local screening of external
electric currents.
\end{enumerate}
 It should now also be clear that the topological
action \pref{top} could have been derived from any of these
conditions. For instance, starting from the condition of local
screening \pref{teom2}, the $BF$ action is obtained simply by
introducing the potential $a$ as a Lagrange multiplier field.

\subsection{The BF-Maxwell theory, Plasmons and Electrodynamic Response}

Thus far we have derived a topological action for the
superconductor which includes the physics of the quasiparticles
and the vortices. There are however, two significant omissions in
this description. The plasmons are missing and so is the defining
characteristic of the superconductor---its electrodynamic
response. As neither of these are topological in nature, this is
sensible. We now show that both of these omissions can be remedied
by keeping the leading irrelevant (but now non-topological) terms
in the action beyond the $BF$ term. These can be guessed on
symmetry grounds alone but to get expressions for their
coefficients we carry out the Gaussian integral over $A$ in
\pref{llow} and obtain the Maxwell-$BF$ Lagrangian, which after
continuation back to Minkowski space becomes, \be{mbf} {\cal
L}_{eff} = \frac 1 \pi \epsilon^{\mu\nu\sigma}b_\mu\pa\nu a_\sigma
-\frac 1 {4e^2}  (f^{(a)}_{\mu\nu})^2 -\frac 1 4 \left(\frac e
{m_s\pi} \right)^2 (f^{(b)}_{\mu\nu})^2
 - a_{\mu}j^{\mu} - b_{\mu}\tilde j^{\mu} \, .
\ee
The equations of motion for the Maxwell-$BF$ theory read,
\be{eommb}
\tilde j^{\mu} &=&
\frac 1 \pi \epsilon^{\mu\nu\sigma}\pa\nu a_\sigma
+  \left(\frac e {m_s\pi} \right)^2 \pa\nu
f_{(b)}^{\nu\mu} \\
j^{\mu} &=&
 \frac 1 \pi \epsilon^{\mu\nu\sigma}\pa\nu b_\sigma
+\frac 1 {e^2}  \pa\nu
f_{(a)}^{\nu\mu} \nonumber \, .
\ee
In the absence of currents, and in Landau gauge ($\partial_\mu
a^\mu = \partial_\mu b^\mu =0$), these can be
combined to give
\be{plasmon}
(\triangle  + m_s^2)a_\mu &=& 0 \nonumber \\
(\triangle + m_s^2)b_\mu &=& 0 \ , \ee which shows that the
spectrum now includes the plasmon modes. We note that an analogous
argument in the quantum Hall problem leads to the
Maxwell-Chern-Simons Lagrangian and thence to the gapped
collective mode\cite{zhang92}.

The reader may wonder at the resemblance of the first of Eqns.
\pref{plasmon} to Eqn. \pref{eom2} with $j^\mu=0$. This is not
coincidental---in going beyond the topological scaling limit we
end up restoring the non-topological parts of the gauge field so
that now $a_\mu$ {\it is} $eA^\mu$ at long wavelengths. This is
also clear from \pref {llow} when we neglect derivative terms.
With this insight we can now confirm that the superconductor is,
in fact, a superconductor. To this end we integrate out $b$ in the
sector without quasiparticles or vortices ($j^\mu = \tilde
j^\mu=0$) to obtain \be{emeff} {\cal L}_{em} = - \frac 1 {4e^2}
(f^{(a)}_{\mu\nu})^2
 - \frac 1 {2e^2} m_s^2 a_\mu a^\mu
\ee which upon variation gives the London equation and thus
superconductivity\footnote{ That an Abelian Higgs model in the
``London limit'' can be rewritten in the dual form, was to our
knowledge first explicitly pointed out by Balachandran and
Teotonio-Sobrinho  who in reference \onlinecite{bala92} considered
the 3+1 dimensional counterparts to Eqns.~\pref{abh2} and
\pref{mbf}.  } Alternatively, we could have explicitly introduced
a background electromagnetic field $\overline A_\mu$ and derived
the London Lagrangian \pref{emeff} directly in $\overline A_\mu$
by integrating out both $a_\mu$ and $b_\mu$.

\section{The ground state degeneracy }

We now return to the analysis of the purely topological field
theory for the low energy excitations of the superconductor. Such
a field theory has no bulk degrees of freedom but will possess
global degrees of freedom which will lead to non-trivial ground
state degeneracies on manifolds of non-trivial topology. In this
section we will first derive the degeneracies predicted by the
$BF$ theory and then understand them physically in the setting of
the \ahm.

As emphasized in the Introduction, one of the hallmarks of a
topologically ordered state is a topology dependent ground state
degeneracy, and a corresponding topological symmetry algebra.
Before analyzing the superconductor it is instructive to recall
how the ground state degeneracy is manifested in the simplest
fractional quantum Hall setting, \ie a
 Laughlin state with filling
fraction $\nu=1/(2k+1)$ on a torus\cite{haldane85}.
In this case the ground state has a $1/\nu$ degeneracy
corresponding to the number of lowest Landau level states for the
center of mass, and the same
degeneracy is obtained from an analysis of the topological
low energy effective action, given  by the
Chern-Simons Lagrangian (\ref{l-cs}).
Here the Wilson loops around the two cycles of the torus form a
canonically conjugate pair,
due to the non-zero commutator $[a_x, a_y]$. The Wilson loops
measure the magnetic fluxes through the holes in the torus,
so it follows that the operators connecting the different
ground states correspond to magnetic flux ``insertions''.

In the superconductor the ground state
degeneracy is again related to the possible values of the Wilson
loops---in this case for the gauge fields $a$ and $b$ appearing in the
topological action. Here, however, there are two conjugate
pairs of variables $(a_x,b_y)$ and $(b_x, a_y)$ so we expect a
squaring of the ground state degeneracy as compared with the
corresponding quantum Hall case. More precisely, the ground
state degeneracy is in both cases determined by the possible ways to
assign commuting fluxes to the ``holes'' in the surface.

\subsection{Ground state degeneracy from the $BF$ theory }
We now formalize this argument, and  show that, in the 2+1
dimensional case, the ground state degeneracy  follows directly
from the $BF$ action \pref{bf} derived in the previous section. We
work on the torus $(L_x,L_y)$.

In the absence of quasiparticles, the $BF$ action can
be written in Hamiltonian form as,
\be{bf-exp}
S = {1 \over \pi} \int d^3 x \{ \epsilon^{ij} \dot a_i b_j +
a_0 ( \epsilon^{ij} \partial_i b_j) +
b_0 ( \epsilon^{ij} \partial_i a_j) \}
\ee
where the Poisson brackets are encoded in the first term,
the Hamiltonian is identically zero,  and
 $a_0$ and $b_0$ are identified as Lagrange multipliers
implementing the constraints,
\be{constraints}
\epsilon^{ij} \partial_i b_j &=& 0 \nonumber \\
\epsilon^{ij} \partial_i a_j &=& 0 \ . \ee On the torus we can
solve these constraints by setting \be{const-soln}
a_i &=& \partial_i \Lambda_a + \bar a^i /L_i \nonumber \\
b_i &=& \partial_i \Lambda_b + \bar b^i /L_i  \ ,
\ee
where $\bar a$ and $\bar b$ are spatially constant, and
$\Lambda_{a/b}$ are periodic functions on the torus.
Upon inserting these forms in the action we find that it
reduces to
\be{tor}
L(\bar a_i,\bar b_i) = \frac 1 {\pi} \epsilon^{ij}
\dot{ \bar a}_i \bar b_j \,
\ee
which identifies $\bar a_i$ and $\bar b_i$ as the physical
degrees of freedom. The remaining, gauge, degrees of freedom
can be eliminated by gauge fixing, \eg by setting
$\partial_i a_i = \partial_i b_i =0$.

From \pref{tor} we obtain the canonical commutation relations,
\be{ccr} [\bar a_x,  \frac 1 \pi  \bar b_y] = i \ \ \ \ \ ; \ \ \
\ \ [\bar a_y, -\frac 1 \pi \bar b_x  ]= i \, . \ee Since these
are two commuting Heisenberg algebras, it naively looks like there
is a continuum of ground states corresponding to different
eigenvalues of \eg $b_x$ and $b_y$. This is however not the case,
since the gauge fields are compact on account of the quantization
of quasiparticle and vortex numbers, as noted previously.
Compactness implies that $\bar a_i \equiv \bar a_i + 2 \pi$ and
$\bar b_i \equiv \bar b_i + 2 \pi$ are angular variables. It
follows that we need instead to consider the operators (Wilson
loops) ${\cal A}_i = e^{i\bar a_i}$ and ${\cal B}_i = e^{i\bar
b_i}$
and their algebras, \\
\be{alg1}
{\cal A}_x {\cal B}_y + {\cal B}_y{\cal A}_x = 0 \ \ \ ; \ \ \
{\cal A}_y {\cal B}_x + {\cal B}_x{\cal A}_y = 0
\, .
\ee
Each of these has a two dimensional representation (via two of
the three Pauli matrices) whence we obtain a $2 \times 2 = 4$-fold
ground state degeneracy on the torus.
It also follows that ${\cal B}_i$ can be
interpreted either as measuring the
$b$-flux or inserting an $a$-flux, and vice versa for
the ${\cal A}_i$.

\subsection{Ground state degeneracy in the abelian Higgs model}
\newcommand{\ket}[1]{|{#1}\rangle }
\newcommand{\bra}[1]{\langle{#1}| }
The above considerations have established a fourfold ground state
degeneracy on the torus (and $4^g$ on genus $g$ surfaces) but left
their physical description obscure. Indeed, the argument beginning
with quasiparticle and vortex braiding is somewhat indirect. To
complete the analysis we now turn to a direct identification of
the states in the \ahm.

The basic observation is our identification of the gauge fields in
the last section. This indicates that in the basis in which ${\cal
A}_i$ are diagonal, the states differ by the amount of magnetic
flux passing through the two holes. At the outset it is important
to emphasize that this is {\it sourceless} flux and better thought
of as the (necessary) assignment of eigenvalues to the Wilson
loops. For ground states, the flux must be an integer multiple of
$\pi$, the superconducting flux quantum. In a theory where the
fundamental charges are $e$, the flux is only defined modulo $2
\pi$, and we get two states for each non-contractible loop. The
operators ${\cal B}_i$ then move the system between these
eigenstates. As the states are degenerate, we can just as well
diagonalize the latter operators and the resulting states are
characterized by even and odd values of the {\it electric} flux.

More explicitly, consider the
position eigenstates $|\phi(\vec r),\vec A(\vec r)\rangle$ of
the gauge and scalar fields in the
Hamiltonian formulation of the abelian Higgs model \pref{abh}.
We can define the action of the operator conjugate to the $x-$Wilson
loop  ${\cal A}_x=\exp(i\oint d x\, A_x) $ on the torus parametrized
by $0 \le x < L_x$ and $0 \le y < L_y$ by
\be{fop}
{\cal B}_y |\phi(\vec r), \vec A(\vec r)\rangle = |e^{i\alpha (\vec r
)}\phi(\vec r),
\vec A(\vec r)+ \frac 1 {2e} \grad \alpha (\vec r)\rangle
\ee
where $\alpha (L_{x},y) = \alpha (0,y) + 2 \pi  $.\footnote{Here and
in the following
we really mean the equivalence class of $\alpha (\vec r)$ under the
addition of functions that are periodic on the torus but we will
be sloppy about this without prejudice to our argument.}
Locally, the effect of the flux insertion operator
${\cal B}_y$ is just a gauge transformation; however, it changes the
sign of the
gauge invariant observable, the Wilson loop ${\cal A}_x$.
This is a global effect, caused by an improper gauge transformation,
that {\it does} change the state. Analogously, we can define the
conjugate pair ${\cal A}_y$, ${\cal B}_x$.

For the pure \ahm\ with only charge-2 matter
we obtain four degenerate states on the torus corresponding to the
possibilities ${\cal A}_i = \pm 1$. Clearly this construction generalizes
to a $4^g$ degeneracy on a closed  surface of genus $g$. As the states
are exactly degenerate, we can just as well choose the basis set to
be eigenstates of the ${\cal B}_i$ instead.

To clarify the meaning of the latter representation it is useful
to give an explicit representation for the operators for the
choice $\alpha(\vec r') = 2\pi\theta(x'-x)$, \be{op}
{\cal A}_x(y) &=& e^{ie \int_0^{L_x} dx'\, A_x(x',y)} \\
{\cal B}_y(x) &=& e^{i \frac{\pi} e \int_0^{L_y} dy'\, E_x(x,y')}
e^{\int d^2r'\,  \alpha(\vec r') \hat \rho(\vec r)} \ee and the
corresponding pair ${\cal B}_x(y)$ and ${\cal A}_y(x)$;
$\hat\rho(\vec r)$ is the charge density operator. Both ${\cal
A}_x(y)$ and ${\cal B}_y(x) $ are clearly gauge invariant, and
have singularities along the lines at $y$ and $x$
respectively.\footnote{ The nature of these singularities are,
however, quite different. The singularity of ${\cal A}_x(y)$
correspond to the creation of a thin line of electric flux, as
discussed in the text, while the singularity in ${\cal B}_y(x) $
is only a gauge artifact. This follows from the relation, $
\bar{\cal B}_y(x_1){\bar{\cal B}}^{-1}_y(x_2) = \exp\left\{
-\frac {in\pi} e \int d^2r'\, \theta(x'-x_1)\theta(x_2-x')
[\partial_{x'} E_x(x',y) - \rho (\vec r')] \right\} \, , $ and
remembering that $\grad\cdot \vec E - \hat\rho$ is the generator
of local gauge transformations. We see that the apparent
singularity at $x$ of the operator $\bar{\cal B}_y(x) $  can be
moved by a regular gauge transformation and thus has no physical
significance.  }

From the canonical equal time commutation relations,
$[A^i(\vec r, t), E^j(\vec r',t)]=i\hbar\delta (\vec r - \vec r')$
and $[\rho(\vec r, t), \phi(\vec r',t)] = i\hbar \phi(\vec r, t)
\delta (\vec r - \vec r')$ follows the commutator algebra,
\be{alg}
{\cal A}_{x(y)} {\cal B}_{y(x)} +
{\cal B}_{y(x)}{\cal A}_{x(y)} = 0 \, .
\ee
which confirms that the operators ${\cal B}_i$ create one
magnetic flux quantum.  We also see that ${\cal B}_{y(x)}$
measure the total electric flux in the $\hat x/\hat y$ direction in units
of $\pi/e$, so that the eigenstates defined by ${\cal B}_i=\pm 1$,
which are symmetric/antisymmetric linear combinations of the magnetic
flux states, have the interpretation of possessing even or odd numbers
of electric flux quanta in the two directions.
Finally, it follows from \pref{alg} that
the Wilson loop ${\cal A}_x(y)$
creates one unit of electric flux in this direction\cite{thooft78} which
completes this dual description.

To  explicitly construct the ground states which all have constant density,
we must include the non trivial winding modes of
the $\varphi$ field,
\be{cmod}
A_{\mu}(\vec r, t) &=&\frac 1 L {\overline A}_{\mu}(t) \\
\varphi (\vec r, t) &=& \varphi_{0}(t) + \frac {2\pi} L \vec n\cdot\vec r
\, , \nonumber
\ee
where $\vec n$ is the winding number vector.
The spatially constant
phase $\varphi_{0}$, conjugate to the total number of particles, can
be absorbed by a spatially constant gauge transformation. The Hamiltonian
in a fixed winding number sector is easily obtained from \pref{abh2}
and given by,
\be{winham}
H_{\vec n} = \half  (\Phi^i_{E})^2 + \frac {m_s^2} 2 ({\overline A}_i +
\frac  \pi e  n_i)^2
\ee
where $\Phi_{E}^i  = L   E^i$ is the spatially constant electric flux
which is conjugate to ${\overline A}_i$,
\be{}
[\Phi_E^i , {\overline A}_j] = i\delta^i_j
\ee
Naively there is a ground state for each winding sector,
and a gap to the plasmon mode at $\hbar m_s$.
Because of gauge invariance
we should however identify all winding number sectors which have the same
value for the Wilson loops ${\cal A}_i = e^{iq \overline A_i}$. For
$q = e$ there are four non-equivalent sectors corresponding to
eigenvalues $\pm 1$ for the operators
${\cal A}_i $. The conjugate operators ${\cal B}_j =
e^{i \frac \pi e \Phi_E}$ are precisely the ``modular electric field"
operators defined by Reznik and Aharonov\cite{rez89}, and ${\cal A}_i $
and ${\cal B}_j$ satisfy the algebra \pref{alg1} which allows us to
identify the potential $b_i$ with the modular electric field.

Three closing comments are in order. \vskip 1mm \noi (1)  A state
with definite ${\cal A}_i$ necessarily has a fluctuating electric
flux present which might seem problematic for a superconductor
which has an infinite conductivity. This is, however, not so. The
crucial point, which is not immediately obvious when one thinks
about classical background electric fields, is that the matter
couples to the vector potential and {\it not} the electric field
and the former clearly has no effect. \vskip 1mm \noi (2) In the
dual states, while there is a definite {\it parity} of the
electric flux, there still isn't an average non-zero flux. Besides,
these states are linear combinations of states that do not possess
a current by the argument in (1). \vskip 1mm \noi (3) Finally, it
is worth emphasizing the importance to our analysis of the
distinction that the gauge potentials $\bar A_{i}$ are not
observables, but the Wilson loops ${\cal A}^q_{i}=e^{iq\bar
A_{i}}$ are, where $qe$ are the charges in the system. Naively, we
would be led to consider states $\ket {\vec n} \equiv
\ket{n_x,n_y}$ with $n_x$ and $n_y$ superconducting flux quanta
through the two holes. However these states are not all distinct
as far as the Wilson loops go and instead form equivalence classes
upon addition of $2/q$ flux quanta in either hole. We have
analyzed the case of the standard superconductor where $q=1$ and
indeed that is true more generally in nature. If however,
fractionally charged matter was present at a fundamental level,
the ground state degeneracies would indeed be different. In such
cases, consistently, the starting topological field theory would
also be different since there would now be a larger set of
braiding phases to encode.

\subsection{Finite size effects and tunneling}
In the last section we were a little sloppy in our discussion
for pedagogical purposes. The ground states that we discussed
arise in two approximations---the neglect of vortex creation/annihalation
in the bulk and in the absence of any other matter, \ie we took the
quasiparticle
gap to be infinity. This had the utility that ground states were
now exactly degenerate for a finite system, but now we can state
the more general situation.

In the general setting we must consider (i) the sensitivity of
the quasiparticle field to the values of  ${\cal A}_{i}$ or
equivalently processes in which two quasiparticles are created
from the vacuum (the condenstate) and then tunnel and recombine
across a non-contractible loop and (ii) a similar process in
which  vortex-antivortex pair is  created from the vacuum
and then tunnels and recombines across a non-contractible loop.
As reviewed in the introduction, such processes are responsible
for motion in the ground state manifold and lead to a lifting
of the topological degeneracy for finite systems.

\begin{figure}
\begin{center}
\includegraphics[width=6in]{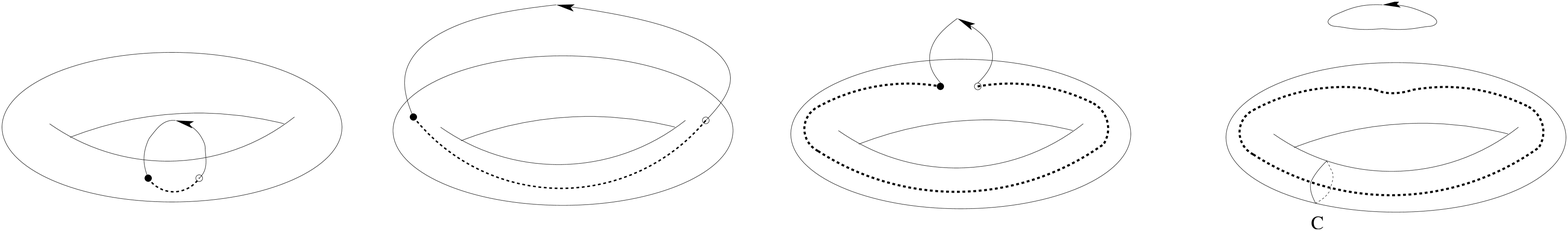}
\caption{
A vortex tunnelling process inserting a unit of magnetic flux inside the torus.
In this visualization it also leaves a flux loop outside, but that is invisible
to the electrons on the surface. This process connects ground states labelled by
opposite values of the Wilson loop $e^{i \oint_C \vec{a} \cdot {\vec dl}}
\equiv e^{i e \Phi_M}$ where $\Phi_M$ is the magnetic flux threading
$C$. }
\label{f3}
\end{center}
\end{figure}

The tunneling process that is easiest to visualize is the
vortex-antivortex tunneling process shown in Fig.~\ref{f3}. Here a unit
vortex-antivortex pair is created, they subsequently move around a
cycle of the torus, and are finally annihilated. During this
process they will insert a unit of magnetic flux inside the torus,
thus changing the value of the corresponding ${\cal A}_{i}$
operator. Thus this process corresponds to a tunneling between the
magnetic flux states, and by itself it will mix them and lift
their degeneracy by an amount $\sim e^{- L_i/\lambda_t}$ where
$L_i$ is the length of the tunneling path, and $\lambda_t$ a
constant of order the screening length. Interested readers can
find a quantitative computation of this process in
Ref.~\onlinecite{vestergren2004}, for the closely related
Fradkin-Shenker system discussed below in Section V-B.

\begin{figure}
\includegraphics[width=5in]{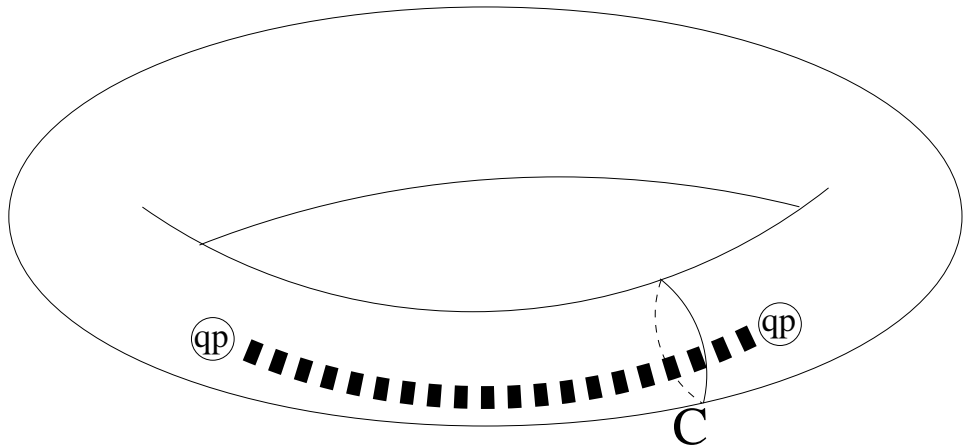}
\caption{
A quasiparticle-pair tunnelling process changing the value of the modular
electric flux, $e^{i \oint_C \vec{b} \cdot {\vec dl}} \equiv
e^{i \frac \pi e \Phi_E}$ where $\Phi_E$ is the surface electric flux crossing
$C$.
}
\label{f4}
\end{figure}

The interpretation of the quasiparticle tunneling process, shown in
Fig.~\ref{f4},
is more subtle. Naively one might think of this as the
charges pulling out an electric flux between them, but since the
superconductor screens, this is not the case on average. What is true instead,
is that a quasiparticle that crosses a surface changes the parity
(evenness/oddness) of the {\it fluctuating} electric flux through its
path. Hence this process connects
the electric flux states and by itself will mix them and lift
their degeneracy by an amount $\sim e^{- L_i/\xi_t}$ where
$\xi_t$ is a constant of order the coherence length.

The actual finite volume ground state in the presence of both
vortex and quasiparticle tunneling will be determined by a
competition between the above two effects.  That the topological
degeneracy is recovered exponentially fast in the linear
dimensions of the system is, as remarked earlier, a hallmark of
topological order.

\subsection{Ground state degeneracy in $d=3+1$}

Finally we present the extension of the discussion in Subsection
A to $d=3+1$. The action \pref{bf3} can be reorganized as
\be{bf3-exp}
S= \frac{1}{\pi} \int d^4 x
\epsilon^{ijk}   \dot a_i b_{jk} + a_0 ( \epsilon^{ijk} \partial_i
b_{jk}) + 2 b_{0i} ( \epsilon^{ijk} \partial_j a_k )  \ ,
\ee
which identifies the four constraints in the problem. As $b_{jk}$
is antisymmetric, its independent components can be identified
as $c^i = \epsilon^{ijk} b_{jk}$ and hence the constraints
rewritten as
\be{bf3-const}
\partial_i c^i &=& 0 \nonumber \\
\epsilon^{ijk} \partial_j a_k &=& 0  \ .
\ee
On the 3-torus, these are solved by setting
\be{bf3-consol}
c^i &=& ( \bar{c}^i +  \epsilon^{ijk}  \partial_j \xi_k  ) / L^{3/2} \nonumber \\
a_k &=& ( \bar{a}_k + \partial_k \Lambda)/ L^{3/2}
\ee
where $\xi$ and $\Lambda$ are periodic functions and we have thus
separated the constant pieces of $c$ and $a$. Upon substituting
these forms in \pref{bf3-exp} we obtain the analog of \pref{tor},
\be{bf3-tor}
L = \frac{1}{\pi} \bar{c}^i \dot{\bar a}_i
\ee
which encodes three commuting Heisenberg algebras and thence, upon
taking account of the compactness of the fields, to $2^3 =8$ states.

\section{Other realizations of the $BF$ theory}

In this section we digress somewhat from the main development
to examine some closely related systems. The systems are
related in that they too are characterized by topological
order described by the $BF$ theory---they all fail to exhibit
local symmetry breaking, a pair of low energy ``matter''
and ``gauge'' excitations with the same braiding phase of $\pi$
and the attendant ground state degeneracy . In a way this is an
example of universality, but in a much more limited sense
than for critical point theories---for the topological
scaling limit keeps much more limited information than
the Wilsonian one. Our examples here are the $Z_2$ lattice gauge
theory, the $U(1)$ lattice gauge theory with charge-2
Higgs scalars, the short ranged resonating valence bond
(RVB) state, and a particular quantum Hall bilayer.

\subsection{$Z_2$ lattice gauge theory}
\newcommand {\zz} {$Z_{2}\ $}

The \zz lattice gauge theory, defined by the Hamiltonian,
\be{zz-h}
H = K  \sum_{\pla_s} \prod_{\langle i j \rangle \epsilon {\pla_s}}
\sigma^z_{ij} + \Gamma \sum_{\langle i j \rangle} \sigma^x_{ij} \, ,
\ee
where the sums are over spatial plaquettes and links,
has been studied extensively, and is well
known to be a topological theory in the $ \Gamma \rightarrow 0$
limit. In this limit all plaquettes must be unfrustrated
in the ground states, $ \prod_{\langle i j \rangle \epsilon {\pla}}
\sigma^z_{ij} =1$.
There are four degenerate ground states on the torus of which
two correspond to the configurations (really their equivalence classes
under local gauge transformations) shown in Fig.~5; the remaining
two are trivial extensions as discussed in the caption.
\begin{figure}
\includegraphics[width=5in]{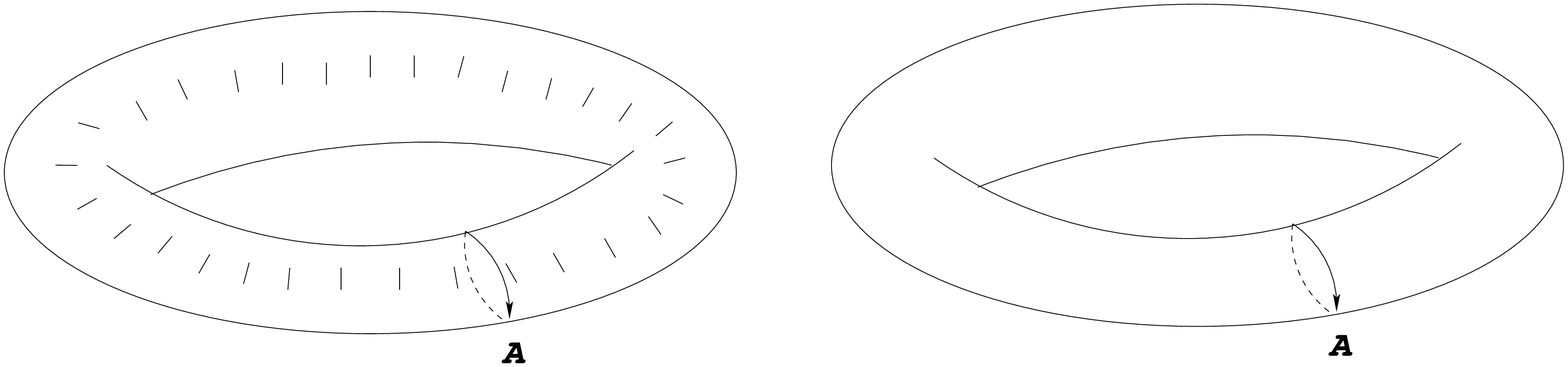}
\caption{ Two of the four ground states of the \zz lattice gauge
theory at zero coupling, on the torus. They differ by the
insertion of a \zz vortex (vison) through one of the holes of the
torus---which is implemented by changing the sign on a string of
bonds as shown. The pair of states thus differ in the sign of the
Wilson loop  $\Pi_C \sigma^z$. The remaining two states differ by
vison insertion in the other hole.} \label{f5}
\end{figure}
Clearly all plaquettes are
nonfrustrated while the Wilson loops around the cycles differ by signs
in the various states. The operators that moves between the different
configurations are singular gauge transformations which are the \zz
counterparts of the $B$ operator introduced in \pref{fop}. The
conjugate, electric field states are discussed \eg in
Ref.~\onlinecite{moes01}.
The excited states of the theory consist of Ising vortices or visons.
If we now couple fundamental Ising matter sources to the gauge
field,
\be{latmat-h}
H_{m}[c] = \beta \sum_{\langle ij \rangle} c_{i}\sigma^z_{ij}c_{j}
\ee
it is easy to see that transporting a ``particle'' around the vison
leads to a $\pi$ phase, \ie  the
\zz gauge theory has the same braiding phases
\cite{sent99,moes01} as the $BF$ theory \pref{top}.
When $K$ is finite but large and the coupling to the matter is weak,
the low energy theory is still the $BF$ theory as we discuss explicitly
next. For variety we will carry out the relevant treatment entirely
on the lattice---it is an interesting feature of this problem that
this can be done.

\subsubsection{The lattice $BF$ action}
As the variables in \pref{zz-h} are discrete, it is most convenient
to work with a discretized time. To this end we begin with the classical
\zz lattice gauge-matter action
\be{zz}
S_{\sigma}[\sigma,c] = - K \sum_\pla
\prod_{\sigma_{ij}\in \pla} \sigma_{ij} -  \beta \sum_{\langle ij \rangle} c_{i}\sigma_{ij}c_{j}
\ .
\ee
where $\sigma_{ij}$ is an Ising variable, and  the sums run
over plaquettes and links on an Euclidian lattice.
We now rewrite \pref{zz} in a  form involving a
lattice version of the $BF$ action, by using  the identity,
\be{iden2}
e^{+K\prod_\pla{\sigma_{ij}} } = f(K) \sum_{\tau =\pm 1}
e^{\tilde\beta\tau}e^{i\frac \pi 4 (1-\tau) (1-\prod_\pla{\sigma_{ij}})} \,
\ee
where $2\tilde\beta = -\ln\tanh K$ and $f(K) = \sqrt{{1 \over 2} \sinh(2K)}$,
for each plaquette, $\pla$, in the  partition
function $Z_{Z_2} = \sum_{\sigma_{ij}} e^{-S_\sigma}$.
This introduces a set of Ising variables,  $\tau_{ij}$ defined
on the links of the dual lattice and, and the partition function can
be expressed as,
\be{zzpart1}
Z_{Z_{2}} = \sum_{\{\sigma_{ij},\tau_{ij}, c_{i} \}} e^{-S_{BF}[\sigma,
\tau ] + \tilde\beta \sum_{\langle ij\rangle} \tau_{ij}
+ \beta \sum_{\langle ij \rangle} c_{i}\sigma_{ij}c_{j} }  \, .
\ee
where,
\be{latbf1}
S_{BF} = -i\frac \pi 4 \sum_{\langle ij\rangle}
(1-\tau_{ij})(1-\prod_{^{\star}\langle ij\rangle }\sigma ) \, .
\ee
Here $^{\star}\langle ij\rangle$ denotes the plaquette on the original
lattice pierced by the link $\tau_{ij}$ on the dual one. Except
for shifts, this term---which multiplies one gauge field $\tau$ with
the flux of the other $\sigma$---is clearly the Ising lattice analog
of the continuum $BF$ term.
This piece of the action was derived by
Senthil and Fisher\cite{sent99}, who also showed
that by partial differentiation it can be
expressed in the alternative form,
\be{latbf2}
S_{BF} = -i\frac \pi 4 \sum_{\langle ij\rangle}
(1-\sigma_{ij})(1-\prod_{^{\star}\langle ij\rangle }\tau ) \ ,
\ee
which is manifestly invariant under the gauge transformation
$\tau_{ij}\rightarrow v_{i}\tau_{ij}v_{j}$ where the $v_i$:s live
on the sites of the dual lattice. Because of this
invariance, we can now recognize \pref{zzpart1} as the restriction
to $v_i=1$ gauge of the manifestly doubly gauge invariant action
\be{zzpart}
Z_{Z_{2}} =
 \sum_{\{\sigma_{ij},\tau_{ij}, c_{i},v_i \}} e^{-S_{BF}[\sigma,
\tau] + \tilde\beta \sum_{\langle ij\rangle} v_i \tau_{ij} v_j
+ \beta \sum_{\langle ij \rangle} c_{i}\sigma_{ij}c_{j} }  \, .`
\ee
If we now specialize to large $K$ (\ie small $\tilde\beta$),
and small $\beta$, we see
that the $BF$ term dominates as promised.

\begin{figure}
\includegraphics[width=5in]{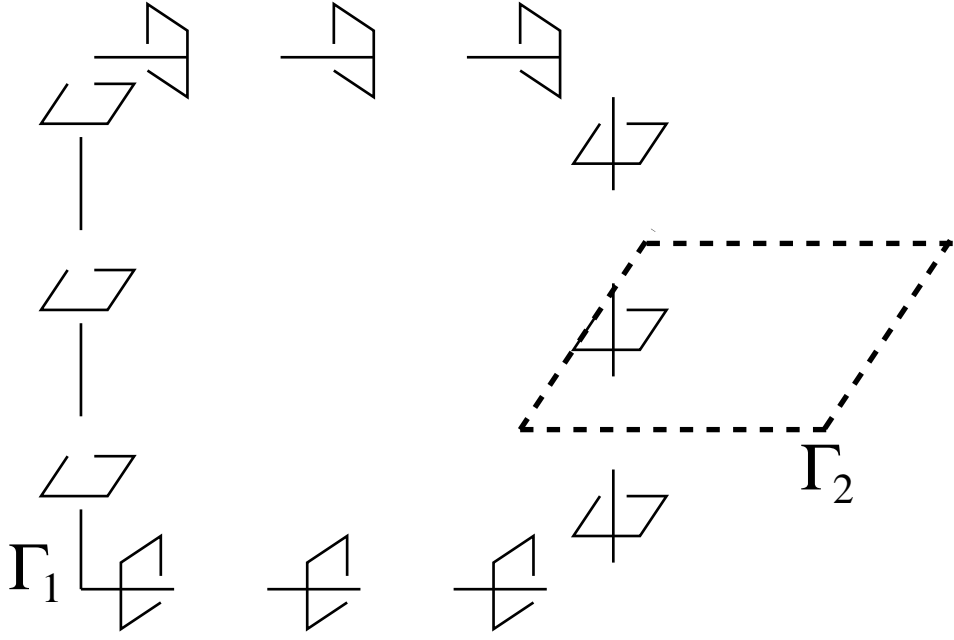}
\caption{ Geometry for establishing the $\pi$ linking phase
between Wilson loops on the direct and dual lattices,
$W_{\Gamma_1}$ and $W_{\Gamma_2}$ in Eqn.~\pref{lloops}}
\label{f6}
\end{figure}

Finally, readers with an appetite for lattice manipulations
can convince themselves that the braiding phases are correctly
reproduced by the lattice $BF$ action by
considering the expectation values of two Wilson loops, one on the
original and one on the dual lattice,
\be{lloops}
\langle W_{\Gamma_{1}}\tilde W_{\Gamma_{2}}\rangle =
\sum_{\{\sigma_{ij},\tau_{ij} \}} e^{-S_{BF}[\sigma,
\tau]} \prod_{\langle ij\rangle\in \Gamma_{1}}\sigma_{ij}
\prod_{\langle kl\rangle \in \tilde\Gamma_{2} } \tau_{kl}
\ee
Expressing $\sigma_{ij} = e^{\frac {i\pi} 2 (1-\sigma_{ij})}$,
and  using  \pref{latbf2} for the action, it is easy to show that
for a link present in the loop $\Gamma_1$, the sum over
 $\sigma_ij =\pm 1$ yields zero if there is not
a  ``wrong sign'' dual plaquette is attached on the dual lattice.
Similarly, for a link not present in $\Gamma_1$, the dual
plaquette must be unfrustrated. As illustrated in Fig.~\ref{f6}, this
implies that a dual loop $\tilde W_{\Gamma_{1}}$ will pick up a
minus sign every time the curve $\Gamma_{2}$ wind around the curve
$\Gamma_{1}$. Clearly the dual of this argument, \ie binding
original plaquettes to the dual links on $\Gamma_{2}$, would give
the same result.

\subsection{$U(1)$ lattice gauge theory with charge-2 Higgs}
\begin{figure}
\includegraphics[width=4in]{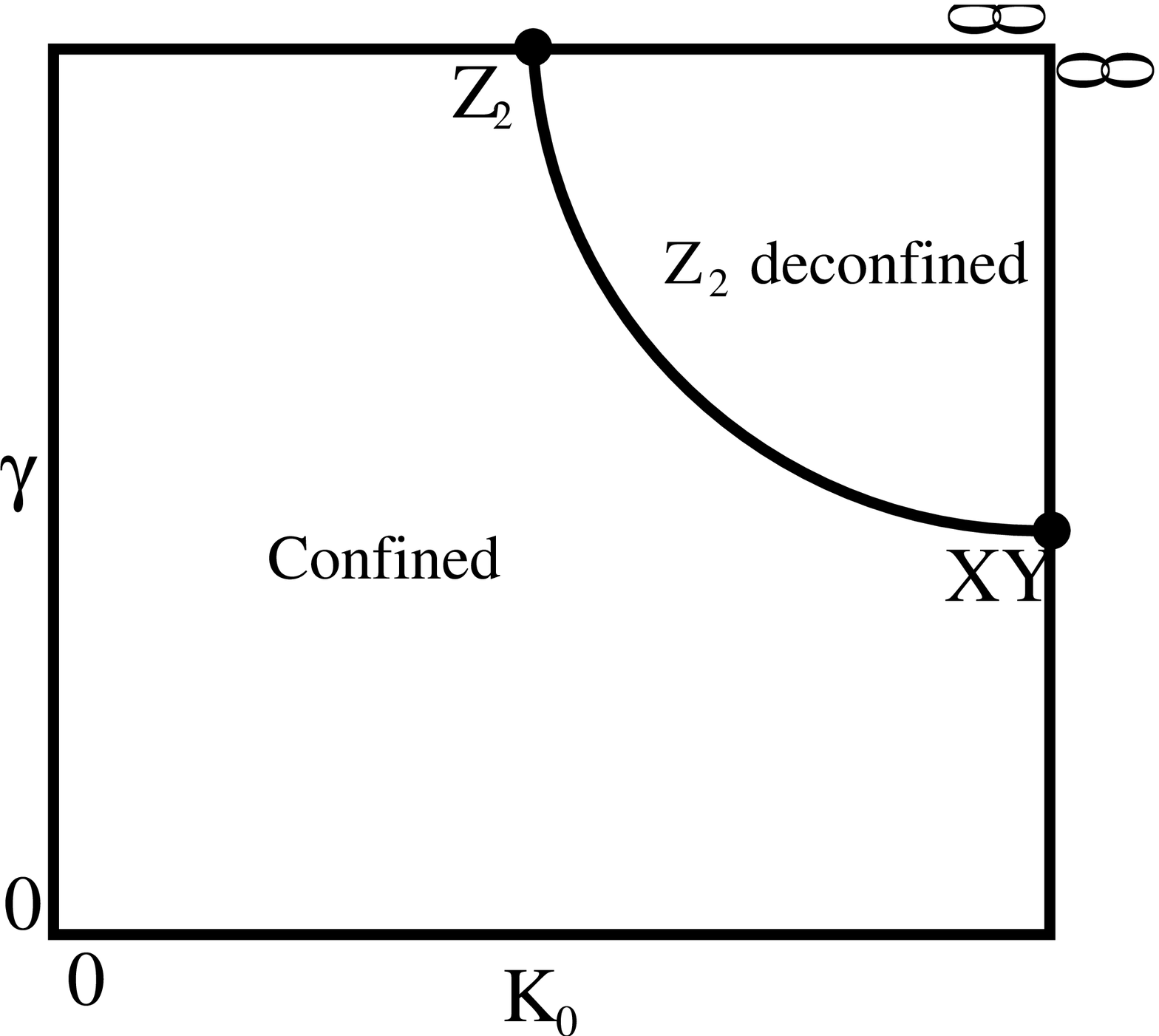}
\caption{
The phase diagram of the $U(1)$ lattice gauge theory with a charge-2 Higgs
scalar in $d=2+1$(after Ref.~\onlinecite{frad79}). In this work we are
concerned with the field theoretic description of the \zz deconfined phase
(upper right portion of the phase diagram). The filled circles on the
boundaries are phase transitions in the universality class of the
indicated models in three spatial dimensions.
}
\label{f7}
\end{figure}

In their influential 1979 paper on gauge-Higgs systems on the
lattice, Fradkin and Shenker \cite{frad79} analyzed a
$U(1)$ lattice gauge theory coupled to charge-2 matter and
showed that it exhibited a phase where the low energy
degrees of freedom reduced to those of the $Z_2$ gauge theory
discussed above, see Fig.~\ref{f7}. Consequently, when the low energy theory is
in its deconfined phase, the gauge-Higgs system is also described
by the $BF$ theory. This system is pretty much a truly lattice
superconductor in that the gauge field also lives on a lattice.
However, the compactness of the {\it microscopic} gauge field
introduces features
that make the characterization of its electromagnetic response
problematic---there seems not to be a definition of the electrical
conductivity that will distinguish the deconfined phase of interest
from the confined phase. This is related to the massive character of
the photon in both phases.
Nevertheless, the model has other uses and
has been extensively invoked in searches for spin liquids
and theories of the cuprates \cite{sent99,moes01} where the
starting problem can often be reformulated as a $U(1)$ theory
coupled to matter but where the gauge field is now generated by
the matter itself and is not related to fundamental electromagnetism.
We now review the  reduction of a lattice superconductor to a
\zz gauge theory by a somewhat different method than used
in the original work.

The starting point is the following lattice action, \be{sclat}
S[U, \Phi] = -\frac {K_0} 2 \sum_\pla \prod_{\langle ij\rangle
\in\pla}       [ U_{ij} + h.c. ]  -  \frac \gamma 2 \sum_{ \langle
ij \rangle } [ \Psi_{i} U_{ij}^{2}\Psi^{\dagger}_{j} + h.c. ] \, .
\ee The gauge potential, $A_{ij}$,  is defined on the links,
$U_{ij} = e^{iA_{ij}}$ and the charge 2 scalar field on sites,
$\Psi_{i} = e^{i\theta_{i}}$, and the two sums are taken over
plaquettes and links of the lattice respectively. Both $A_{ij}$
and $\theta_i$ are angular variables defined on the interval
$[0,2\pi]$, and in terms of these the action takes the form,
\be{sclat2} S[A, \theta] = -  K _0\sum_{\pla}  \cos (F_{\pla}) -
\gamma \sum_{ \langle ij \rangle} \cos [\theta_{i} -\theta_{j} -
2A_{ij} ]\, , \ee where $F_\pla =F_{ijkl} = A_{ij} + A_{jk} -
A_{lk} - A_{il} $, is  the lattice field strength of the plaquette
$\pla  = (ijkl)$.

What is of relevance here is that on the $K_0 = \infty $ line in
the phase diagram, Fig.~\ref{f7},  the theory \pref{sclat} becomes
a \zz gauge theory. To show this, we make the following
decomposition of the gauge potential, \be{dec} A_{ij} = \half
[a_{ij} + \pi(1-\sigma_{ij})] \, , \ee corresponding to $U_{ij} =
\sigma_{ij}e^{\frac i 2 a_{ij}}$, where $\sigma_{ij}$ is an Ising
variable, and the range of the angular variable $a_{ij}$ is again
from 0 to $2\pi$. We then use the following identity, \be{iden}
\int_{0}^{2\pi} \frac {d\phi} {2\pi} f(\phi) = \half \sum_{\sigma
= \pm 1} \int_{0}^{2\pi} \frac {d\chi} {2\pi} f(\half[\chi +
\pi(1-\sigma)]) \ee to rewrite the partition function as,
\be{lpart} Z = \prod_{\langle ij \rangle, k} \int_{0}^{2\pi}
dA_{ij}d\theta_{k} \, e^{-S[A_{ij}, \Phi_{k}]} = \prod_{\langle ij
\rangle, k } \int_{0}^{2\pi} da_{ij}d\theta_{k} \, \half
\prod_{\langle lm \rangle }\sum_{\sigma_{lm} = \pm 1 }
e^{-S^{'}[a_{ij}, \theta_{k}, \sigma_{lm}]} \, . \ee In the
$\gamma \rightarrow \infty$ limit it is convenient to use a
unitary gauge where $\theta_i = 0$ and the action for $a_{\mu}$
takes the form, \be{aact} S[a, \sigma] =- K_0 \sum_{\pla} \cos
(\half f_{\pla} ) - \prod_{\langle ij\rangle \in\pla} \sigma_{ij}
- \gamma \sum_{\langle ij\rangle} \cos (a_{ij}) \, , \ee with
$f_{\pla}$ is the lattice field strength corresponding to
$a_{ij}$. The effective \zz action is now defined as, \be{zzdef}
e^{-S_{\sigma}[\sigma]} = \prod_{\langle ij \rangle}
\int_{0}^{2\pi} da_{ij} e^{-S^{'}[a, \sigma]} \, . \ee and can be
computed in a perturbative expansion in $1/\gamma$. To lowest
nontrivial order we obtain, \be{latgauge} S_{\sigma}[\sigma, c] =
-  K  \sum_{\langle ij\rangle \in\pla} \prod_\pla \sigma_{ij} \ee
where $ K = K_0(1 + \frac 1 {4\gamma} )$. We now add a charge
$q=1$ field $\phi_i = e^{i\vartheta_i}$ with the action $S[U,\phi]
=( \beta_0/2) \sum_{\langle ij\rangle } [\phi_i
U_{ij}\phi_j^\dagger + h.c.] $. Decomposing the angular variable
as $\vartheta_i = \half (\xi_i +\pi c_i)$  we have the identity,
$\cos (\vartheta_i - \vartheta_j - A_{ij}) = c_{i}\sigma_{ij}c_{j}
\cos\half (\xi_i - \xi_j - a_{ij})$, and integrating $a_{ij}$ and
$\xi_i$, gives the action \be{latmat} S_{m}[\beta, c] = -\beta
\sum_{\langle ij \rangle} c_{i}\sigma_{ij}c_{j} \ee which
describes the coupling of an Ising matter field. Combining
\pref{latgauge} and \pref{latmat} we regain the \zz lattice action
\pref{zz}, and hence, by the results of the previous subsection,
 the $BF$ theory as the low energy, purely topological
description of the compact lattice superconductor.

\subsection{RVB State}

The short ranged RVB state of a quantum Heisenberg magnet, first
proposed by Anderson \cite{pwa-rvb}, is a liquid of spins paired
into local singlets. In the extreme short ranged case the
wavefunction is made up solely of configurations $| c \rangle $ in
which each spin is paired with exactly one nearest neighbor spin.
A prototypical liquid wavefunction is then an equal amplitude
superposition
\begin{equation} \label{eq:rvbstate}
|\psi \rangle =  \sum_c | c \rangle
\end{equation}
of such configurations. The physics of the nearest
neighbor problem is captured in the quantum dimer model \cite{qdm}
and following the demonstration that the triangular lattice
quantum dimer model supports a liquid phase \cite{rvbtrilat} it
has become clear that this generalizes to other non-bipartite
lattices.

\begin{figure}
\includegraphics[width=5in]{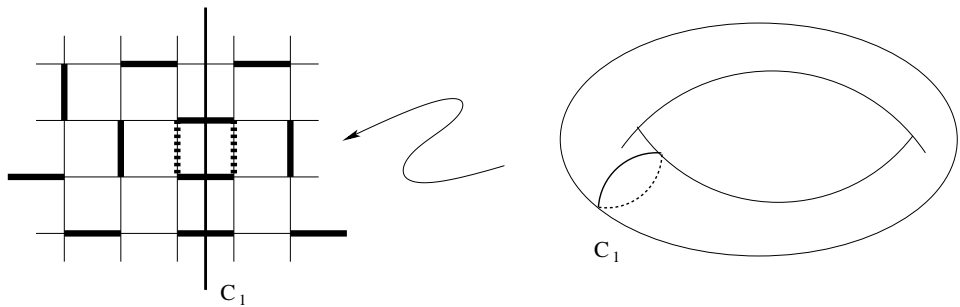}
\caption{ Topology of dimer coverings: The number of dimers
crossing the non-contractible loop $C_1$ can only change by an
even number under a local dimer dynamics, \eg the resonance move
shown by the dashed lines changes the number by two. Consequently,
the ground states of the quantum dimer model on the torus can be
labelled, in the deconfined phase, by the  number of dimers {\it
modulo} $2$ crossing the non-contractible loops. } \label{f8}
\end{figure}

This liquid, RVB, phase can be readily seen to lead to a $4^g$
ground state degeneracy \cite{readchak}. As shown in Fig.~\ref{f8},
the parity of the number of dimers crossing a non-contractible
loop is invariant under a local dimer dynamics which thus yields
two distinct liquid states for each such loop. In terms of our
previous discussion for superconductors this is the analog of
the parity of the electric flux.

\begin{figure}
\includegraphics[width=5in]{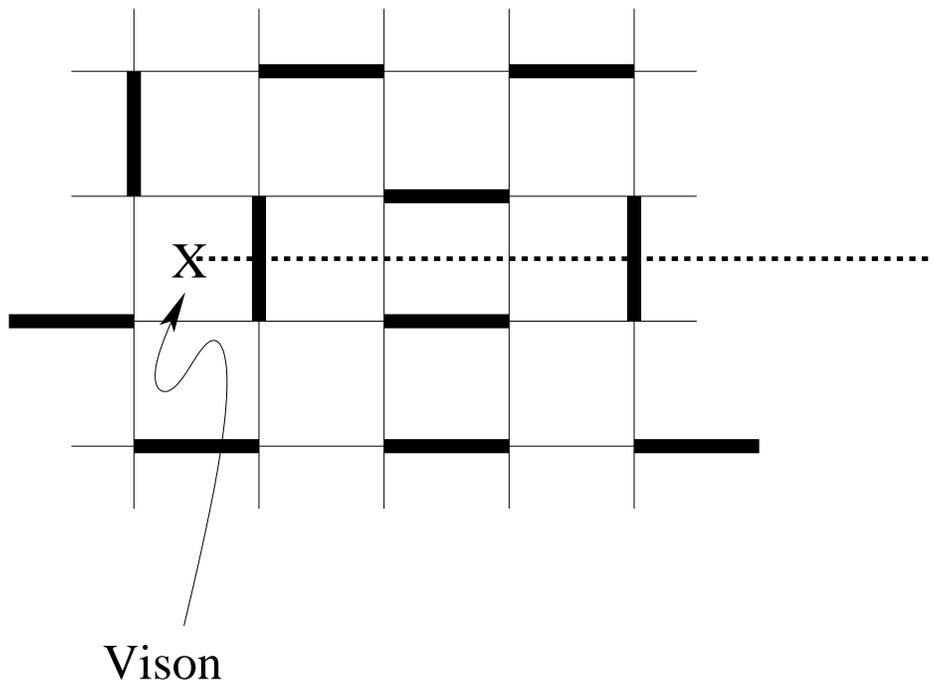}
\caption{The vison involves a string going out to infinity. A
dimer configuration $c \rangle$ is now weighted by $(-1)^{N_s(c)}$,
where $N_s(c)$ is the number of dimers crossing the string.
}
\label{f9}
\end{figure}

The excitations of the RVB state are spinons and visons
(vortices). A spinon is an unpaired spin while a vison involves a
phase string (Fig.~\ref{f9}). It is not difficult to see that these
gapped excitations have the familiar topological interaction with
a mutual braiding phase factor of $-1$ arises. It is also an
instructive exercise to see that the tunneling of spinons and
vortices leads to the lifting of the ground state degeneracy. From
all of this it follows then that the RVB state again has a
topological description by the $BF$ action.

While the pictures drawn above pertain to two dimensions, recently
it has been shown that the quantum dimer model on the FC lattice
exhibits an RVB phase \cite{ms3drvb} which is then characterized
by the $3+1$ dimensional version of the $BF$ action. Finally, we
should note that in the case of the RVB, the microscopic problem
is that of a strongly coupled gauge theory so a trivial reduction
to the topological actions is not feasible, as it was for the
weakly coupled phases of the \zz gauge theory discussed above.

\subsection{A quantum Hall interpretation of the $BF$ theory}

Finally, we observe that the $BF$ theory can be taken to describe
a somewhat unusual quantum Hall system.

According to Wen and Zee\cite{wenzee92} the general form of the
topological action for an abelian quantum Hall liquid is (in an
obvious form notation), \be{hac} {\cal L}_{qh} = \frac 1 {4\pi}
K_{IJ}a^{I}da^{J} + \frac e {2\pi} t_I Ada^I - a^{I}j^{I} \ee
where $K_{IJ}$ is a symmetric matrix , and $t_{I}= \delta_{I1}$ a
vector, both with integer entries. This action leads to a ground
state degeneracy $|{\rm det}K|^{g}$ on a surface of genus $g$ and
the true electrical charge, $q$, of a quasiparticle with charges
$l_I$ with respect to the gauge fields $a^I$ is given by $q =
-et_I K^{-1}_{IJ}l_J$.

For quantum Hall systems the matrix $K$ is taken to be positive
semidominant, corresponding to the lack of  time reversal
invariance. The formalism can be extended, however, to time
reversal invariant systems by expanding the allowed $K$ matrices.
In our case $K^{IJ} = 2\sigma^{x}_{IJ}$ reproduces the $d=2$ BF
action. As a check, on a torus $|{\rm det}K|^{g} = 4^{1} = 4$ as
derived before.

An alternative quantum Hall representation is obtained by the
transformation,
\be{qhtr}
a^{1} \equiv a = R + L \\
a^{2} \equiv b = R - L \nonumber
\ee
giving
\be{qhac2}
{\cal L}_{qh} = \frac 1 \pi (RdR - LdL) + j(R+L) + \tilde j (R-L)
\ee
\ie two decoupled $\theta = \pi/4$ liquids with {\it opposite}
sense of time reversal breaking. Note that although the
elementary quasiparticles acquire a $e^{i\pi/4}$
phase under exchange, the original charges and vortices carry charge
with respect to both layers (or, equivalently, can be thought of as
composites of charges in the two layers). It is an elementary
exercise to verify that the combined Berry and exchange phases come
out correctly if we restrict ourselves to this sector of the
expanded problem.

\subsection{Instantons and the nature of charge}

In this section we have covered a diverse set of systems that
give to the $BF$ theory in their topological scaling limit. Evidently,
as we move away from that limit the differences among the systems
will reassert themselves. Here we wish to comment on one of these
differences, namely the nature of the charges in the various
systems.

We note that the $BF$ theory formally involves a $U(1)$ gauge field
and hence a coupling to $U(1)$ currents. But this is misleading
since in writing it we have really only encoded a finite amount
of information on braiding phases---in particular these phases
are insensitive to whether the quasiparticle and vortex currents
are truly conserved or only conserved modulo 2. Among the systems
we have considered, both currents are integer valued for the
hypothetical quantum Hall system. In the ordinary superconductor
the vortex number is integer valued but quasiparticle number is
only defined modulo 2 since a pair of quasiparticles can always
disappear into the condensate.
In the \zz gauge theory, both currents are evidently only defined
modulo 2 and since the Fradkin-Shenker problem reduces to the
former the same is true there.

These differences are, of course, built into the microscopic
actions. Of interest here is how they can be incorporated
in the $U(1)$ description as we move beyond the topological
scaling limit.
The solution to this puzzle is that compact gauge fields
permit finite action instantons that break the corresponding
$U(1)$ down to $Z_2$.

For the $a$ field the instantons are unit strength monopoles
that can create or destroy two Abrikosov
flux lines of strength $1/2e$, as illustrated in Fig.~\ref{f8}.
The strength of the tunneling will depend on microscopic
details, which determines the magnitude of the instanton action.

\begin{figure}
\includegraphics[width=5in]{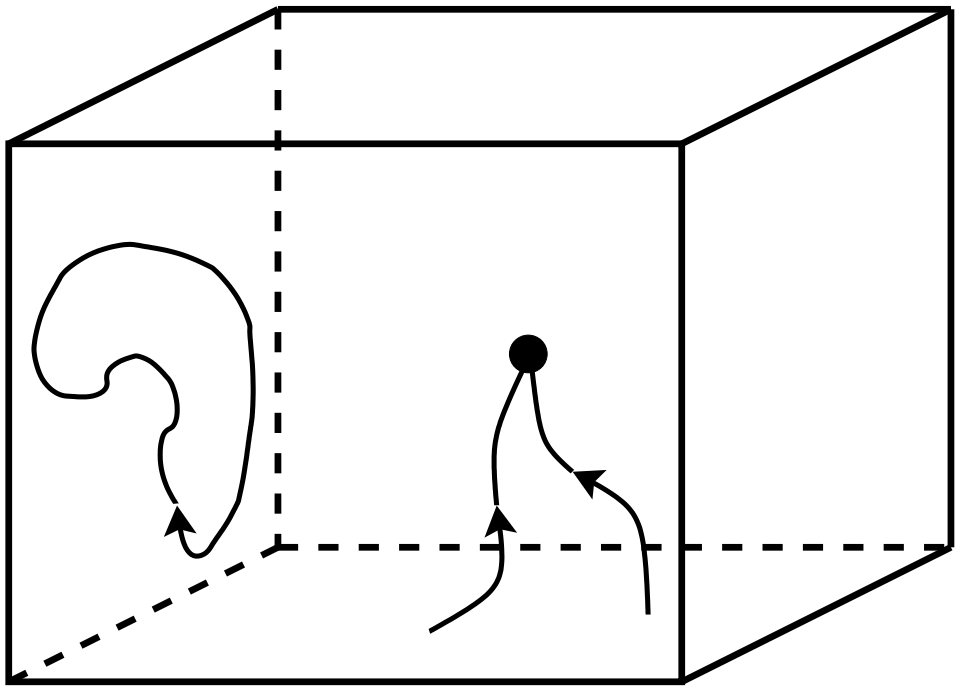}\caption{
Virtual vortex-antivortex fluctuations represented as a space-time vortex loop.
Also shown are two vortices annihilating on a monopole.
}
\label{f10}
\end{figure}

We now also learn how to incorporate the charge non-conserving
effects of Cooper pair breaking and formation in the context of
$BF$ theory  - it simply amounts to allowing monopole
configurations in the dual gauge field $b$! It is an interesting
technical challenge to actually derive this prescription directly
from the path integral formulation of the full abelian Higgs
model.

Returning to our original question it is now clear that the
inclusion of instantons is the mechanism by which the
different conservation scenarios are distinguished beyond
the topological scaling limit. The quantum Hall realization
includes none, the ordinary superconductor includes (on
reasonable scales!) only the $b$ monopoles and the \zz
gauge theory and the Fradkin-Shenker problem require both
$a$ and $b$ monopoles.

\section{Edge States}

Returning to the topological order characteristics for quantum
Hall states listed in the introduction, we see that we have found
analogs of all of them in superconductors save one---these are
``edge states'' to which we now turn. The existence of edge
states, \ie degrees of freedom localized near the boundary of a
manifold with a boundary, can be deduced quite generally. To begin
with, one can see qualitatively that fractionalization in the bulk
implies that the missing fractional quantum numbers of the
quasiparticles must migrate to the boundary and thence that the
boundary must support degrees of freedom capable of absorbing
these quantum numbers. This can be sharpened once one has a
topological field theory in hand. While on closed manifolds the
topological field theory has only global degrees of freedom, in
the presence of a boundary it ceases to be purely topological and
now exhibits boundary degrees of freedom.

From the quantum Hall effect we however know that the details of
the boundary theory is, in general,
not coded in the bulk topological action,
but depends crucially on the nature of the confining potential.
For instance, a polarized Laughlin state with a sharp edge
will have a single chiral edge mode with a velocity given
by the $\vec E\times \vec B$ drift at the edge. In a softer
potential the edge can reconstruct giving pairs of
counterpropagating modes which in general develop a gap.

With suitable boundary conditions, the topological field theory
does define the phase
space of a minimal theory needed for current conservation.
In the quantum Hall case it is the electric current of the
bulk quantum liquid and its associated quasiparticles. In
the case of the superconductor there are two currents,
described by the gaugefields $a_{\mu}$ and $b_{\mu}$
corresponding to charge and vorticity respectively.
Thus, from the knowledge of the quasiparticles in
the bulk one obtains a listing of the different sectors of the
edge theory---which correspond to the independent ways in which
quasiparticles in the bulk can influence the edge dynamics. This
further allows identification of the operator spectrum at the edge.
What remains is the identification of the edge Hamiltonian and while
that can be constrained on symmetry grounds there remain details
that only microscopics can fill in.

The choice of boundary conditions for the topological field theory
is crucial - different choices give different dynamics, or even no
dynamics at all. In the quantum Hall case the boundary conditions
are well understood, at lest in the simplest cases, but to our
knowledge there is no rigorous derivation from a microscopic
approach. A brief review of the quantum Hall case is given in the
Appendix. In the case of the superconductor the situation is  less
clear. A microscopic approach would be to study  \eg the abelian
Higgs model \pref{abh} in the presence of a interface, carefully
follow the steps leading to the topological $BF$ action \pref{top}
and deduce the relevant boundary conditions, which would depend on
the nature of the interface. We shall not take this route but
rather, in the spirit of Section V-D, assume the kind of boundary
conditions used to analyze immunocompetent quantum Hall systems. A
discussion of different boundary conditions in $BF$ theories and
the abelian Higgs model can be found in the work of Balachandran
\etal\cite{bala92,bala94}.

\subsection{$BF$ theory on manifold with boundary in $d=2$}

As briefly explained in the Appendix, the pertinent starting point
is the Hamiltonian form \pref{bf-exp} of the $BF$ action, which
we now consider on a manifold
$\Omega$ with a boundary $\partial \Omega$ parametrized by $x_1$.
Under the boundary conditions
$a^{0}|_{\partial\Omega} = b^{0}|_{\partial\Omega} =0$, this
action coincides with the covariant expression \pref{top} restricted
to the same domain.
The constraints \pref{constraints} are now solved by
\be{const-soln-d}
a_i &=& \frac 1{2} \partial_i \Lambda_a \nonumber \\
b_i &=& \frac 1{2} \partial_i \Lambda_b \ ,
\ee
where $\Lambda_{a/b}$ take arbitrary values at the boundary.
Upon inserting these forms in the action we find that it
reduces to
\be{baction-adb}
S = -{1 \over 4 \pi } \int_{\partial\Omega}  d^2 x\,
\partial_0 \Lambda_a \, \partial_1 \Lambda_b
\ee
which shows that the only degrees of freedom live at the edge and
that their phase space is that of a one dimensional boson with
both chiralities present---from \pref{baction-adb} we can
read off the canonical commutation relations,
\be{ccr-edge}
[\Lambda_a(x,t), - \frac{1}{4 \pi} \partial_1\Lambda_b(y,t)] =
i \delta(x-y)  \ .
\ee

The analysis thus far is modified when there are quasiparticles
and/or vortices present in the bulk. Now the boundary line
integrals of the gauge fields are non-zero but quantized, so it is
necessary to allows the edge bosons to wind along the edge. Their
winding numbers, \be{wnos}
N_a &=& \frac 1 {2 \pi} \int_{\partial\Omega} dx^1 \partial_1 \Lambda_a  \nonumber \\
N_b &=& \frac 1 {2 \pi} \int_{\partial\Omega} dx^1 \partial_1 \Lambda_b
\ee
count the numbers of vortices and quasiparticles in the bulk respectively.
Equivalently, they count the screening charges at the boundary so we can identify
the edge vortex and quasiparticle densities as
$\frac 1 {2\pi} \partial_1 \Lambda_a$ and
$\frac 1 {2\pi} \partial_1 \Lambda_b$ respectively.
In turn this identifies $\psi_a^\dagger \sim e^{-i \Lambda_b/2}$ as the edge
vortex
creation operator and $\psi_b^\dagger \sim e^{-i \Lambda_a/2}$ as the edge
quasiparticle creation operator while
Cooper pairs (valence bonds) and $2 \pi$ vortices are created by
$e^{-i \Lambda_a}$ and $e^{-i \Lambda_b}$ respectively.
It is not hard to see that in a sector
with $N_{a/b}$ odd $\psi_{b/a}^\dagger$ picks up a factor of $-1$
upon circling the edge and hence exhibits the correct braiding.
Finally, one technical point is worthy of note. The quantization conditions
\pref{wnos} and the set of operators identified here are {\it not}
those of a compact boson of any specified radius. While this is important
for a detailed understanding of the spectrum, it will not matter for the
rest of our discussion.\footnote{As we were finishing this paper there
appeared Ref.~\onlinecite{Gukov:2004id} which also notes this point
as a special case in the course of a more general analysis of $BF$ theories
in $2+1$ dimensions. Their point of departure, however, could not be more
different!}

We turn now to the Hamiltonian, where the true nature of the currents, discussed
in the last section, becomes important. If $a$ and $b$ are truly $U(1)$ fields
then the edge Hamiltonian must conserve vortex and quasiparticle number and
we conclude that it takes the form
\be{bf-edgeham1}
 H = \int_{\partial \Omega} dx^1 \left(
\frac {v_1} 2 (\pa x\Lambda_a)^2 + \frac {v_2} 2 (\pa x \Lambda_b)^2 \right) \, ,
\ee
plus higher gradient corrections. The quadratic cross-term is
ruled out by time reversal invariance. In this case the edge is gapless
and exhibits Luttinger liquid behavior.

For all our remaining cases however both charges are not conserved. At
a minimum Cooper pair creation/annihilation is allowed so that we must add the
term
\be{cpcreate}
 H_a = \int_{\partial \Omega} dx^1
\frac{g_a}{2} ( e^{-i \Lambda_a} + e^{i \Lambda_a} ) \equiv
g_a \cos(\Lambda_a)
\ee
to ${\cal H}$. For the Fradkin-Shenker problem, the RVB state and the
$Z_2$ gauge theory we also need to add the dual process of vortex pair
creation/annihilation,
\be{vcreate}
H_b = \int_{\partial \Omega} dx^1
\frac{g_b}{2} ( e^{-i \Lambda_b} + e^{i \Lambda_b} ) \equiv
g_b \cos(\Lambda_b) \ .
\ee
The resulting theory  $H + H_a + H_b$ is a dual sine-Gordon model with
one of the cosines being
generically the most relevant operator. It follows then, that the edge is
generically gapped.

\subsection{$BF$ theory on manifolds with boundary in $d=3$}

We now return to the action \pref{bf3-exp} and the constraints \pref{bf3-const}
but now on a manifold with a boundary. In the line with the discussion in
$d=2$ we now write the solution to the constraints as
\be{bf3-console}
c^i &=& \epsilon^{ijk}  \partial_j \xi_k / L^{3/2} \nonumber \\
a_k &=& \partial_k \Lambda  / L^{3/2}
\ee
where the boundary values of $\xi_k$ and $ \Lambda$ are now unconstrained.
The action now takes the form
\be{bf3-eact}
S &=& \frac{1}{\pi} \int_{\partial \Omega} d^3 x
\dot \Lambda (\epsilon^{ijk} \partial_j \xi_k)_n \nonumber \\
&=&  \frac{1}{\pi} \int_{\partial \Omega} d^3 x
\dot \Lambda ( \partial_1 \xi_2 - \partial_2 \xi_1 )
\ee
where on the first line the subscript $n$ indicates the normal to the bounding
surface and on the second we have taken the latter to have local coordinates
$(1,2)$. Evidently this is the symplectic structure of a scalar field with
$\frac{1}{\pi}(\partial_1 \xi_2 - \partial_2 \xi_1)$ playing the role of the
conjugate momentum,
\be{ccr-edge3}
[\Lambda({\bf x},t), \frac{1}{\pi} (\partial_1 \xi_2 - \partial_2 \xi_1)
({\bf y},t)] = i \delta({\bf x-y})  \ .
\ee
Unlike in $d=2+1$ where the two edge fields enter symmetrically, we
see that they have different character in $d=3+1$.

The analysis of sectors is more complex for this reason. The
presence of quasiparticles in the bulk leads to the quantization
\be{wnos31}
N_ \xi = \frac {-1}{\pi} \int_{\partial\Omega} dx^1 dx^2
(\partial_1 \xi_2 - \partial_2 \xi_1) \ .
\ee
With vortex lines first consider the situation of the infinite cylinder.
Here the line integral
\be{wnos-vort}
N_a &=& \frac {-1} { \pi} \int_{\partial\Omega} dx^1 \partial_1 \Lambda
\ee
around the circumference will count the number of vortex lines running
parallel to the cylinder axis. Sectors with $N_a \ne 0$ are manifestly
locally stable but they are at infinite energies relative to the ground
state. For generic bounded geometries, the situation is more
complicated: vortex lines in the bulk will have to exit the surface at
two points which then define vortices in the field $\Lambda$. While
one can formally define sectors of the edge theory with an arbitrary
number of such vortex/anti-vortex pairs, since the bulk dynamics will
force the vortex lines to move about, the actual problem can no longer
 be studied purely at the edge, so   a
edge/bulk separation is no longer possible .

Turning now to the edge ``vertex'' operators, we note that the
quasiparticle creation operator $\psi_\xi^\dagger \sim e^{i\Lambda}$
increases $N_\xi$ by one. The existence of such a local operator is
to be expected, e.g. in the RVB problem one can see that a spinon
created in the bulk leads to the creation of a spinon at the boundary
and the latter is equally a local object. For vortex lines let us
restrict ourselves to the case of the cylinder. Here $\psi_a^\dagger
\sim e^{i \pi_0 \frac{2 \pi R x}{L}}$ generates shifts between
different values of $N_a$ where $\pi_0 = \frac{1}{L} \int dx^2
(\partial_1 \xi_2 - \partial_2 \xi_1)$. This operator is non-local,
again as one expects.

The conserving Hamiltonian is now
\be{bf-edgeham31}
 H = \int_{\partial \Omega} dx^1 dx^2 \left(
\frac {v_1} 2 (\nabla \Lambda)^2 + \frac {v_2} 2 (\nabla
(\partial_1 \xi_2 - \partial_2 \xi_1))^2 \right)  \, ,
\ee
and the addition of quasiparticle and vortex line creation/annihilation
again generically  gives rise to gaps.

\subsection{Gapless edges and topology changing phase transitions}

As we have noted above, except in the case where both quasiparticle
and vortex currents are truly $U(1)$ currents, the edges will be
gapped for generic values of the coupling constants. There are special
values of the couplings, however, for which the edges are gapless.
This gaplessness arises because in both $d=2$ and $d=3$ the two
perturbations break the $U(1)$ symmetry down to $Z_2$ in dual ways
and an Ising transition separates the two phases obtained when
just one of perturbations dominates. In $d=2$ this is well understood
to happen along the line $g_a = g_b$ and the resulting theory
is the familiar Majorana fermion of the critical Ising model. In
$d=3$, while an exact solution is evidently not feasible, general
symmetry arguments again indicate that the critical theory is
that of the Ising model.

There are two settings in which the critical Ising theory arises
naturally in $BF$ systems. First, it arises on a single edge if
the microscopic model has an additional symmetry. Such a lattice
model in $d=2$ has been constructed by Wen \cite{wen-majorana}
and exhibits a gapless
Majorana fermion at the edge---we direct the reader there for
the details. Currently we do not know of a model of a continuum
superconductor that has this feature.

\begin{figure}
\includegraphics[width=5in]{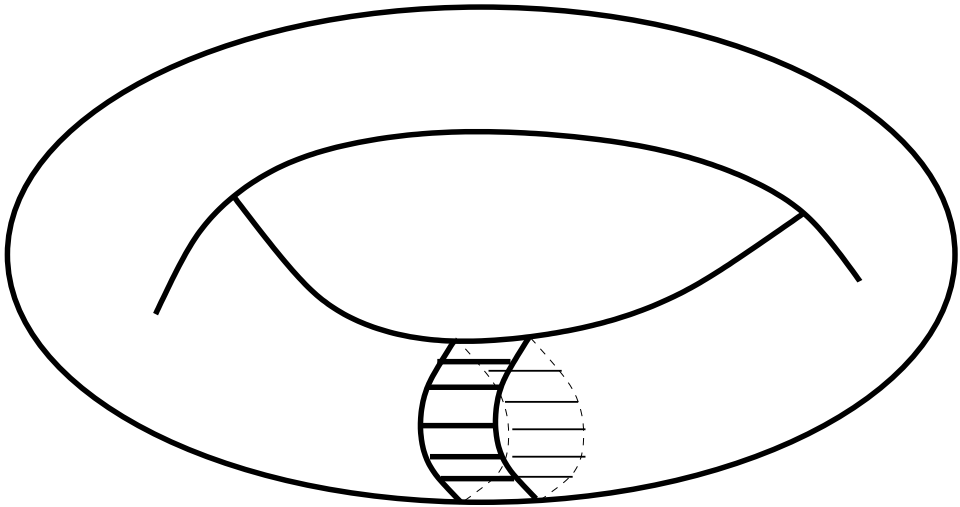}
\caption{ Weakening the indicated row of plaquettes produces as
set of low energy edge states (Section VI). At a critical value of
these couplings a ``topology changing phase transition'' ensues. }
\label{f11}
\end{figure}

The second setting is that of the ``topology changing
phase transition'' first discussed by Wen and Niu \cite{wen&niu}
in the context of the
quantum Hall effect and then by Senthil and Fisher \cite{sf-tcpt}
in their
investigation of $Z_2$ gauge theories of correlated systems---they
are also responsible for the nomenclature.
Here the idea is that we construct a closed manifold by sewing up
a manifold with two boundaries, for specificity consider taking
a cylinder and sewing it up into a torus (Fig.~\ref{f11}). The ground state
degeneracies before and after sewing are different,  so as
a function of the strength of the coupling there must be
a phase transition along the way. In the $BF$ problem the disconnected
edges are gapped and hence the cylinder exhibits a two-fold
degeneracy from the one closed, non-contractible loop. The fully
connected edges must give rise to a further two-fold degeneracy
and hence we may expect an Ising transition
{\it en route}.\footnote{As an aside we note that for the $\nu=1/2$
bosonic quantum Hall state the two fold degeneracy is reached from
a phase with gapless edges and hence the transition should be
expected to be of the Kosteritz-Thouless type, as shown in \cite{wen&niu}.
This will also be the case in the quantum Hall bilayer of Section
V-D.}
For the $Z_2$ gauge theory, this can be seen explicitly \cite{sf-tcpt}
by tuning the strength of a line of plaquettes. For superconductors
the details are not readily worked out but the general arguments
apply just as well.

To round out this discussion, we now review how the topology
changing phase transition appears from the perspective of the
$BF$ theory. Returning to our favorite Lagrangian \pref{bf-exp}
and parametrizing the cylinder with $(x,y)$, $y$ periodic,
we now write,
\be{}
a_1 &=& \partial_1 \Lambda_a \nonumber \\
a_2 &=& \bar a^2 /L_2 + \partial_2 \Lambda_a
\ee
and
\be{}
b_1 &=& \partial_1 \Lambda_b \nonumber \\
b_2 &=& \bar b^2 /L_2 + \partial_2 \Lambda_b
\ee
where $\Lambda_{a/b}$ are periodic in $x_2$ alone. These lead
to the Lagrangian
\be{cylinder}
\pi L = \frac{\bar b_2}{L_2} \int dx^2 (\dot\Lambda_{au} - \dot\Lambda_{al})
+ \frac{\bar a_2}{L_2} \int dx^2 (\dot\Lambda_{bu} - \dot\Lambda_{bl})
+  \int dx^2 (\dot\Lambda_{au} \partial_2 \Lambda_{bu} -
            \dot\Lambda_{al} \partial_2 \Lambda_{bl})
\ , \ee which exhibits the symplectic structure of the two bosons
on the upper ($u$) and lower ($l$) edges. The addition of
quasiparticle/vortex pair creation on each edge will then gap
both.

Bringing the edges together will generate couplings between them by
tunneling processes involving Cooper pairs and pairs of vortices.
Quasiparticle tunneling will be a higher energy process while fractional
vortices cannot tunnel across a gap---the same argument excludes
quasiparticle tunneling in the quantum Hall effect version of this
problem. Generically these processes will not be of equal strength
and so the sewing of the torus will proceed in stages. For concreteness
let us take the Josephson coupling to be the larger of the two. The
corresponding term
$$
\cos(\Lambda_{bu} - \Lambda_{bl})
$$
will drive an Ising transition past which it will set
$$
\Lambda_{b} = \frac{{\bar b_1} x_1}{L_1} + \Lambda_b^\prime \ ,
$$
and hence reduce \pref{cylinder} to
\be{}
\pi L = \frac{\bar b_2}{L_2} \int dx^2 (\dot\Lambda_{au} - \dot\Lambda_{al})
+ \frac{\bar a_2}{\dot{\bar b_1}}
+ \int dx^2 \partial_2 \Lambda_b^\prime (\dot\Lambda_{au}  - \dot\Lambda_{al})
\ ,
\ee
which is now the theory of a single boson running along the cut. The
growth of the remaining coupling will freeze
$\dot\Lambda_{au}  - \dot\Lambda_{al}$ and via a second Ising transition
will lead to the purely topological action \pref{baction-adb}. This is
the transition studied in \cite{sf-tcpt}.

\section{Related work}

The fundamental observation at the heart of our work is the
topological interaction between superconducting vortices and
quasiparticles. This has a venerable history in the condensed
matter literature---it is present in the solution of the
Bogoliubov-de Gennes equations in the presence of a vortex where
one sees a half-integer shift in angular momentum for
quasiparticle states well beyond the penetration depth or
coherence length \cite{dGbook}. Its modern formulation by
Goldhaber and Kivelson \cite{goldkiv} built on the analysis of
quasiparticle fractionalization by Kivelson and
Rokhsar\cite{kiv90} referred to in the introduction as well on the
earlier work of Reznik and Aharanov \cite{rez89}. In the high
energy theory literature this interaction is central to the
``discrete gauge theory'' work starting with that of Krauss and
Wilczek \cite{kra89} reviewed in Ref.~\onlinecite{propitius}.

The particular formulation used in this paper, that of topological order,
briefly appeared in Wen's early work \cite{wen91} and with a
comment on the excitation spectrum that overlooked the dual role of the
quasiparticles. Interestingly, Balachandran and collaborators \cite{bala92}
considered the problem of writing a topological field theory for the
superconductor in 3+1 dimensions, as well the implications for edge
structure. While our conclusions, independently reached, about the $BF$
action in $d=3+1$ are identical, our discussion of edge structure is quite
different and {\it prima facie} somewhat disconnected from the concerns in the
earlier work. The ground state degeneracy and its lifting by tunneling
are themes missing from this prior work.

Finally it is also worth repeating that the topological order discussed
here for the superconducting state with electromagnetic interactions
is different from the topological order discussed for states obtained
by disordering an uncharged superconductor \cite{itp}. While the mathematics
is similar, the physical meanings of the gauge fields are quite different.

\section {Summary and Closing Remarks}

In this paper we have revisited the notion of ordering in a gapped
superconductor. We find that the low energy, topological, physics
of such superconductors fits conveniently into the paradigm of
topological order exemplified by quantum Hall states.
Mathematically, the topological $BF$ action captures this physics
in all dimensions and we have used that to discuss ground state
degeneracies and edge structure. Keeping the leading operators
beyond the topological limit recovers the more familiar
electrodynamics of the superconducting state. We have also
examined physically distinct systems, such as the short ranged RVB
state, which share the same topological field theory and can be
considered members of a ``topological universality class". There
are two obvious directions in which this analysis can be extended.
First, gapless superconductors with gapless quasiparticles can be
given a low energy description by the action \be{[} {\cal L} =
\frac 1 \pi \epsilon^{\mu\nu\sigma}b_\mu\pa\nu a_\sigma -
b_{\mu}\tilde j^{\mu}  - a_{\mu}j^{\mu} + {\cal L}_{\rm qp} \, .
\ee which generalizes \pref{top} by keeping the dynamics $ {\cal
L}_{\rm qp}$ of the gapless quasiparticle current $j^\mu$. This is
no longer a purely topological action but we expect that its
detailed analysis will capture the low energy physics of gapless
superconductors \cite{ohswip}. It is also interesting to explore
the connection between this formulation and the ``quantum order''
idea of Wen \cite{wen-qorder}, who has proposed that the
projective construction of interacting quantum states from
mean-field states is a way to classify them. For superconductors,
the mean field state can be taken to be the standard neutral BCS
state tensored with the classical state in which the
electromagnetic field given by the London equation  ${\bf A}=
-\lambda {\bf J}$. A projection enforcing Gauss's law will then
yield a state that presumably has the correct physics of the
combined matter-field system. Such a construction can accommodate
both gapped and gapless superconducting states.

Second, as in the work on the Hall effect, our abelian analysis
suggests the prospect of finding ``non-abelian superconductors''
or ``non-abelian RVB states'' whose physics is captured by
non-abelian generalizations of the abelian $BF$ theory. This could
proceed via the non-abelian $BF$ theory discussed in the
literature or (in $d=2+1$) by the quantum Hall bilayer
construction discussed in Section V-D. We note that the latter
possibility has also been out forward by Freedman \etal
\cite{freedman03} from a point of departure very different from
our own but with the same effect of accommodating $P$ and $T$
invariant states within the Chern-Simons class of topological
field theories. We also note that Higgs phases of non-abelian
Yang-Mills theories are known to exhibit topological interactions
based on discrete non-abelian groups \cite{propitius} and there is
also a condensed matter construction of such discrete non-abelian
gauge theories based on Josephson junction arrays
\cite{doucotioffe}. An analogous survey of these systems from the
topological order viewpoint could well prove useful.

\acknowledgements
We would like to thank Eduardo Fradkin, Xiao-Gang Wen and Paul
Fendley for instructive discussions. THH thanks the Swedish Research
Council for support and Princeton University's Department of Physics
for its hospitality. SLS and VO are grateful to the
National Science Foundation (grant NSF-DMR-0213706) and the David and
Lucile Packard Foundation for support.

\appendix
\section{The quantum Hall system with a boundary}

Here we review the derivation of the edge action for abelian quantum Hall
states specializing, for simplicity, to the Laughlin fractions $\nu
= 1/k$. We begin with the topological action and
include background gauge fields that allow computation of the
electromagnetic response.
 Also including a quasiparticle current, $j$, we have the following dual
Chern-Simons theory,
\be{bulk}
{\cal L} =  \frac k {4\pi} ada  +\frac e {2\pi} adA - ja \, .
\ee
Integrating out the $a$ field we get
\be{emres}
{\cal L} =  -\frac {e^{2}}  {4\pi k} AdA + \frac{e}{k} jA -
\frac{\pi}{k}  j \frac 1 d j
\, ,
\ee
where the first term gives the quantum Hall conductance, $\sigma_{H} =
\nu e^{2}/h$, the second shows
that the quasiparticles have charge $\nu e$, and the third encodes
the statistical interaction making them $\theta = \nu \pi$ anyons.
For the following analysis we shall take $j=0$.

On a closed surface of genus $g$ the analysis of \pref{bulk}
proceeds along the lines discussed in Section IV and yields a
Hilbert state of $k^g$ states which are the degenerate ground
states of the quantum Hall fluid. We now consider how the analysis
proceeds for a bounded region $\Omega$ that has a one dimensional
boundary $\partial \Omega$---the edge of the system.

A proper specification requires that we pick a boundary condition
\cite{elitzur,wenrev}, and, as discussed in the text, this should
follow from the microscopic physics. In the present case, there
are several ways to establish that the boundary supports a gapless
chiral edge mode. We now show how this feature is reproduced by
taking $a_0 =0$ at the boundary.\cite{wenrev}

\subsection{The edge action}
With this choice, and the absence of background fields, the action
corresponding to \pref{bulk} can be reorganized as, \be{a00} S =
\frac k {4\pi} \int_\Omega d^3 x \left[ a_2\dot a_1 - a_1\dot a_2
+ 2a_0b \right] \, . \ee to exhibit $a_0$ as a Lagrange multiplier
field that imposes the constraint $b  = 0 $. This can be solved as
\be{console} a_j = -{1 \over k} \partial_j \chi \ee and on
substituting this back in \pref{a00} we find that \be{baction} S =
-{1 \over 4 \pi k} \int_{\partial\Omega}  d^2 x\, \partial_0 \chi
\partial_1 \chi
\ee
where we have chosen to parametrize the edge by the co-ordinate labelled
1.

We see, consequently, that for a bounded region the action depends
only upon the field $\chi$ at the boundary, \ie the only physical
degrees of freedom live at the boundary. The remaining degrees of
freedom are purely gauge ones and should be eliminated by a
suitable choice of gauge for the $a$ field. Further, we see that
the physical degrees of freedom are those of a chiral boson since
the action \pref{baction} specifies the canonical commutation
relations of such a boson. The connection to microscopics is
transparent for a circular droplet in symmetric gauge---the
excitations have only momentum of only one sign. Absent an edge
confining potential, these states can be thought of as degenerate
ground states as indeed they appear in our choice of a theory with
a vanishing Hamiltonian when $a_0 =0$. For the alternate,
chirality breaking, boundary condition, $a_0 + v a_1 =0$ the same
analysis yields the nonvanishing Hamiltonian $H= {v \over 4 \pi k}
\int_{\partial\Omega}  d^2 x\, (\partial_1 \chi)^2$. Alternatively
we keep the boundary condition $a_{0}=0$ and just add the above
Hamiltonian as an allowed term in an effective edge action.

\subsubsection{Including background gauge fields}

If we now consider the response of the system to background (external)
electromagnetic fields $A_\mu$, we are led to the background gauge
invariant action (we set $e=1$),
\be{bulk-b}
S[a,A] =  \frac{k}{4\pi} \int_{\Omega}  d^3 x\, \left[
ada   + \frac{2}{k}adA \right]
 \ \ \ \ \ ,
\ee
which can be rewritten in the equivalent form
\be{bulk-b2}
S[a,A] =  \frac{k}{4\pi} \int_{\Omega}  d^3 x\, \left[ada
+ \frac{2}{k} Ada \right]
+  \frac{1}{2\pi k} \int_{\partial\Omega}  d^2 x\, \left[
 A_0 a_1 -  A_1 a_0 \right]
\ee
from which it is easy to see, by functional differentiation with
respect to the background field, that we have coupled the latter
to the bulk current
\be{bulkcurr}
j^\mu_{\rm bulk} = - \frac{1}{2 \pi} \epsilon^{\mu \nu \lambda}
\partial_\nu a_\lambda
\ee
and the edge current
\be{edgecurr}
j^0_{\rm edge} = - \frac{1}{2 \pi} a_1 \  \ , \ \  j^1_{\rm edge} =
\frac{1}{2 \pi} a_0 \ .
\ee

We can now analyze this action with the same boundary conditions
on the $a$ fields, \ie $a_0 =0$. Then,
\be{a02}
S[a,A] = \frac k {4\pi} \int_\Omega d^3 x
\epsilon^{ij} \dot a_i a_j  + 2 a_0 (\epsilon^{ij} \partial_i a_j
+ {1 \over k} \epsilon^{ij} \partial_i A_j) + {2 \over k} (
 \epsilon^{ij} \dot A_i a_j + \epsilon^{ij} a_i \partial_j A_0)
\, .
\ee
The constraint now takes the form $\epsilon^{ij} \partial_i (
a_j + {1 \over k} A_j) =0$ which has the solution
\be{consoln2}
a_j = -{1 \over k} (A_j + \partial_j \chi) \ .
\ee
To maintain background gauge invariance we require that
$\chi \rightarrow \chi - \Lambda$ when $A_j \rightarrow
A_j + \partial_j \Lambda$.

Substituting this back in \pref{a02} we find that it reduces
to
\be{aedgef}
S = - \frac 1 {4\pi k} \int_\Omega d^3 r \epsilon^{\mu \nu \lambda}
A_\mu \partial_\nu A_\lambda -
\frac 1 {4\pi k} \int_{\partial\Omega}  d^2 r\, D_0 \chi D_1 \chi
+ A_0 \partial_1 \chi - A_1 \partial_0 \chi
\ee
which yields both the bulk electrodynamics response captured
in the Chern-Simons term and the coupling of the edge degree
of freedom to the background field. In the above, $D_{0/1} \equiv
\partial_{0/1} - a_{0/1}$. One can check directly
that the above form is background gauge invariant and that the
equation of continuity of current is obeyed at the boundary
when the edge current is included. \ie the anomaly cancels.
Again, for the
alternate boundary condition, $a_0 + v a_1 =0$ the same analysis
adds the gauged Hamiltonian $H= {v \over 4 \pi k}
\int_{\partial\Omega}  d^2 r\, (D_1 \chi)^2$.

Let us finally comment on background fields in the case of the
superconductor. When electromagnetism is dynamical, a background
field can only be introduced as a technical device to calculate
current correlation functions. In some models of strongly
correlated 2d electron systems there are electrically charged
particles coupled to {\it bona fide} 2D gauge fields. For example
we could consider {\it holons} obtained by removing the electron
from a site occupied by and RVB spinon. In this case one can
introduce a background electromagnetic field as in the quantum
Hall case and calculate response functions. Also, a background
field corresponding to $b_\mu$ can be introduced as a technical
device to calculate vortex current correlation functions.


\begin{thebibliography}{99}

\bibitem{chluben} P. M. Chaikin and T. L. Lubensky, {\sl Principles
of Condensed Matter Physics}, Cambridge, 1995.

\bibitem{anderson84} See, \eg P. W. Anderson,
{\em Basic Notions of Condensed Matter
Physics}, Benjamin/Cummings, 1984.

\bibitem{compete} See  K. McElroy \etal, cond-mat/0404005 and
references therein.

\bibitem{wencsl} X.-G. Wen, Phys. Rev. {\bf B40}, 7387 (1989);
Int. J. Mod. Phys. B {\bf 4}, 239 (1990).

\bibitem{wen&niu} X.-G. Wen and Q. Niu, Phys. Rev. {\bf B41},
9377 (1990).

\bibitem{wen91} X.-G. Wen, Int. J. Mod. Phys. B {\bf 5}, 1641
(1991).

\bibitem{wenrev} X.-G. Wen, Advances in Physics {\bf 44}, 405 (1995).

\bibitem{gmop} One non-local operator with algebraic correlations
was identified by S. M. Girvin and A. H. MacDonald, Phys. Rev.
Lett. {\bf 58}, 1252 (1987) and used to exactly reformulate the
dynamics by S.-C. Zhang, T. H. Hansson and S. Kivelson,
Phys. Rev. Lett. {\bf 62}, 82 (1989). A second such operator due to
N. Read, Phys. Rev. Lett. {\bf 62}, 86 (1989) actually condenses and
has been given a field theoretic implementation in
R. Rajaraman and S. L. Sondhi, Int. J. Mod. Phys. B {\bf 10}, 793 (1996).


\bibitem{msf2001} R. Moessner, S. L. Sondhi and E. Fradkin Phys. Rev. {\bf B65},
024504 (2001).

\bibitem{elitzur} S. Elitzur, Phys. Rev. {\bf D12},  3978 (1975).

\bibitem{kiv90} S. A. Kivelson and D. S. Rokhsar, Phys. Rev. {\bf B41},
11693 (1990).

\bibitem{frad79} E. Fradkin and S. H. Shenker, Phys. Rev. {\bf D19}, 3682
(1979).

\bibitem{dirac} P. A. M. Dirac, Can. J. Phys. {\bf 33}, 650 (1955).

\bibitem{kk85} T. Kennedy and C. King, Phys. Rev. Lett. {\bf 55}, 776
(1985).

\bibitem{fradsussdual} E. Fradkin and L. Susskind,
Phys. Rev. {\bf D17}, 2637 (1978).

\bibitem{thooft78} G. 't Hooft, Nucl. Phys. {\bf B138}, 1, (1978);
{\em ibid}  {\bf B153}, 141, (1979).


\bibitem{fm01} J. Fr{\"o}hlich and P. A. Marchetti, Phys. Rev. {\bf D64},
014505 (2001).

\bibitem{weinbII} See \eg S. Weinberg, {\em The Quantum Thory of
Fields}, vol II,
Cambridge University Press (1996).

\bibitem{stringbreak} B. Andersson, G. Gustafson,
G. Ingelman and T. Sj{\"o}strand, Phys. Rep. {\bf 97}, 31 (1983).

\bibitem{pnoz} See \eg chapt. 3 in D. Pines and P. Nozieres, {\em
The Theory of Quantum Liquids}, Addison-Wesely, 1989.

\bibitem{itp} L. Balents, M. P. A. Fisher and C. Nayak,
Int. J Mod. Phys. B {\bf 12}, 1033 (1998).

\bibitem{goldkiv}
A.~S.~Goldhaber and S.~A.~Kivelson, Phys.\ Lett.\ B {\bf 255}, 445
(1991).

\bibitem{swieca76} J. A. Swieca, Phys. Rev. {\bf D13}, 312 (1976).

\bibitem{buchholz79} D. Buchholz and K. Fredenhagen, Nucl. Phys. {\bf
B154}, 226, (1979).

\bibitem{frohlich-top} J. Frohlich and T. Kerler, Nucl. Phys. B {\bf 354}, 369 (1991).

\bibitem{dGbook} P. G. de Gennes, {\em Superconductivity of Metals and Alloys},
Addison-Wesley, 1989.

\bibitem{kra89} L. M. Krauss and F. Wilczek, Phys. Rev. Lett.
{\bf 62}, 1221 (1989).

\bibitem{rez89} B. Reznik and Y. Aharonov, Phys. Rev. {\bf D40}, 4178
(1989).

\bibitem{blau} M. Blau and G. Thompson, Ann. Phys. {\bf 205}, 130 (1991).

\bibitem{berg95} M. Bergeron, G. W. Semenoff and R. J. Szabo,
Nucl. Phys. {\bf B437}, 695, (1995).

\bibitem{bala92} A. P. Balachandran and P. Teotonio-Sobrinho, Int. J. Mod. Phys., {\bf A8}, 723 (1993),
       hep-th 9205116.

\bibitem{bard78} K. Bardakci and S. Samuel, Phys. Rev. {\bf D18}, 2849
(1978).

\bibitem{zhang92} S. C. Zhang, Int. Journ. Mod. Phys. B, {\bf 6},
25 (1992).

\bibitem{haldane85} F. D. M. Haldane, Phys. Rev. {\bf B64}, 2529 (1985).

\bibitem{vestergren2004} A. Vestergren, J. Lidmar and T. H. Hansson,
cond-mat/0402566.

\bibitem{moes01} R. Moessner, S. L. Sondhi and E. Fradkin Phys. Rev. {\bf B65},
024504 (2001).

\bibitem{sent99} T. Senthil and M. P. A. Fisher,, Phys. Rev. {\bf B62},
7850 (1999).

\bibitem{pwa-rvb} P. W.  Anderson, Mat. Res. Bull. {\bf 8}, 153 (1973).

\bibitem{qdm} D. S. Rokhsar and S. A.  Kivelson, \prl {\bf 61}, 2376 (1988).

\bibitem{rvbtrilat} R. Moessner and S. L. Sondhi, \prl {\bf 86}, 1881 (2001).

\bibitem{readchak} N. Read and B. Chakraborty, Phys. Rev. {\bf B40}, 7133 (1989);
X. G. Wen, Phys. Rev. {\bf B44}, 2664 (1991).

\bibitem{ms3drvb} R. Moessner and S. L. Sondhi, Phys. Rev. {\bf B68},
184512 (2003).

\bibitem{wenzee92} X. G. Wen and A. Zee, Nucl. Phys.
{\bf B15}, 135 (1990).


\bibitem{bala94} A. P. Balachandran, L. Chandar and E. Ercolessi,
hep/th 9411164.

\bibitem{Gukov:2004id}
S.~Gukov, E.~Martinec, G.~Moore and A.~Strominger,
arXiv:hep-th/0403225.

\bibitem{wen-majorana} X.-G. Wen, Phys. Rev. Lett. {\bf 90}, 016803 (2003).

\bibitem{sf-tcpt} T. Senthil and M.P.A. Fisher, Phys. Rev. {\bf B63}, 134521 (2001)

\bibitem{propitius}
M.~de Wild Propitius and F.~A.~Bais, arXiv:hep-th/9511201.

\bibitem{ohswip} T. H. Hansson, V. Oganesyan and S. L. Sondhi,
work in progress.

\bibitem{wen-qorder} X.-G. Wen, Phys. Rev. {\bf B65}, 165113 (2002).

\bibitem{freedman03} M. Freedman \etal, cond-mat/0307511.

\bibitem{doucotioffe}B. Doucot, L.B. Ioffe and J. Vidal,
cond-mat/0302104.

\end{thebibliography}
\end{document}